\documentclass[aps,a4paper,amsfonts,amsmath,amssymb]{revtex4}
\usepackage{graphicx}
\begin{document}
\title{The Impact of Heterogeneous Trading Rules on the Limit
Order Book and Order Flows}
\author{Carl Chiarella}
\email{carl.chiarella@uts.edu.au}
\affiliation{School of Finance and Economics University of
Technology, Sidney, \\ PO Box 123, Broadway NSW 2007 Australia}
\author{Giulia Iori}
\email{g.iori@city.ac.uk}
\affiliation{Department of Economics, City University,\\
Northampton Square London, EC1V 0HB, UK}
\author{Josep Perell\'o}
\email{josep.perello@ub.edu}
\affiliation{Departament de F\'{\i}sica Fonamental, Universitat de Barcelona\\
Diagonal, 647 Barcelona E-08028, Spain}
\date{\today}

\begin{abstract}
In this paper we develop a model of an order-driven market where
traders set bids and asks and post market or limit orders according
to exogenously fixed rules. Agents are assumed to have three
components to the expectation of future asset returns,
namely-fundamentalist, chartist and noise trader. Furthermore agents
differ in the characteristics describing these components, such as
time horizon, risk aversion and the weights given to the various
components. The model developed here extends a great deal of earlier
literature in that the order submissions of agents are determined by
utility maximisation, rather than the mechanical unit order size
that is commonly assumed. In this way the order flow is better
related to the ongoing evolution of the market. For the given market
structure we analyze the impact of the three components of the
trading strategies on the statistical properties of prices and order
flows and observe that it is the chartist strategy that is mainly
responsible of the fat tails and clustering in the artificial price
data generated by the model. The paper provides further evidence
that large price changes are likely to be generated by the presence
of large gaps in the book.
%\noindent{\bf KEYWORDS: Market microstructure, limit orders, fundamentalism, chartism, large fluctuations.}\\
%\noindent{\bf JEL: C6,D4, G1}
\end{abstract}

\maketitle

\section{Introduction}

Traders willing to trade in electronic markets can either place market orders, which are immediately executed at the current best listed price, or they can place limit orders. Limit orders are stored in the exchange's book and executed using time priority at a given price and price priority across prices. A transaction occurs when a market order hits the quote on the opposite side of the market.

In the last few years several order driven microscopic models have been introduced to explain the statistical properties of asset prices, we cite in particular Bottazzi et al. (2005), Consiglio et al. (2003), Chiarella and Iori (2002), Daniels et al. (2002), Li Calzi and Pellizzari (2003), Luckock (2003), Raberto et al. (2001)
and Gil-Bazo et al. (2005). See also Slanina (2007) for a recent overview.

The aim of Chiarella and Iori (2002) was to introduce a simple
auction market model in order to gain some insights into how the
placement of limit orders contributes to the price formation
mechanism. The impact on the market of three different trading
strategies: noise trading, fundamentalism and chartism were
analyzed. It was shown that the presence of chartism plays a key
role in generating realistic looking dynamics, such as volatility
clustering, persistent trading volume, positive cross-correlation
between volatility and trading volume, and volatility and bid-ask
spread.

While several microstructure models have been able to generate price trajectories that share common statistical properties with real asset prices, few studies have so far investigated the properties of generated order flows and of the order book itself. Recently, the empirical analysis of limit order data has revealed a number of intriguing features in the dynamics of placement and execution of limit orders. In particular, Zovko and Farmer (2002) found a fat-tailed distribution of limit order placement from the current
bid/ask (with an exponent around $1.49$). Bouchaud et al. (2002) and Potters and Bouchaud (2003) found a fat-tailed distribution of limit order arrival (with an exponent roughly equal to $0.6$, smaller than
the value observed by Zovko and Farmer) and a fat-tailed distribution of the number of orders stored in the order book (with exponent of about $0.3$).

In order to build a model that can incorporate these recent
empirical findings of limit order data, we here extend the original
model of Chiarella and Iori (2002) in two main respects. First,
agents apply different time horizons to the different components of
their strategies; longer time horizons for the fundamentalist
component and shorter horizons for chartist and noise trader
component. Second, agents are not constrained to merely submit
orders of size one, but rather submit orders given by an asset
demand function determined in the traditional economic framework of
expected utility maximisation. In this way the ongoing evolution of
the market feeds back into the asset demands of the different
agents. We simulate our model and compare its qualitative
predictions for different strategies of our population of traders.
We show in particular with the introduction of chartist strategies
into a population of utility optimizing traders that our model is
not only capable of generating realistic asset prices, fat tails and
volatility clustering but, is also capable of reproducing the
empirically observed regularities of order flows.

The analysis of order book data has also added to the debate on what
causes fat tailed fluctuations in asset prices. Gabaix et al. (2003)
put forward the proposition that large price movements are caused by
large order volumes. A variety of studies have made clear that the
mean market impact is an increasing function of the order size, in
contrast Farmer et al. (2004) have shown that large price changes in
response to large orders are very rare. Furthermore these authors
have also shown that an order submission typically results in a
large price change when a large gap is present between the best
price and the price at the next best quote (see also Weber and
Rosenow (2005) and Gillemot et al. (2005)). We show in section 3
that in the model proposed in this paper large returns are also
mostly associated with large gaps in the book.

The paper is structured as follows; in section \ref{TheModel} we model the fundamentalist, chartist and noise trader components of expectations and the way in which agents form their demands for the risky asset. In section \ref{sim anal} we undertake a number of simulations of the model under our choice of the asset demand function to determine how well the empirical facts referred to earlier are reproduced. Section \ref{conclude} concludes and suggests some avenues for future research. The appendices contain a number of technical derivations.

\section{The Model}\label{TheModel}

We assume that all agents know the fundamental value $p^f_t$ of the
asset, which we take to follow a geometric Brownian motion. Agents
also know the past history of prices. At any time $t$ the price is
given by the price at which a transaction, if any, occurs. If no new
transaction occurs, a proxy for the price is given by the average of
the quoted ask $a^q_t$ (the lowest ask listed in the book) and the
quoted bid $b^q_t$ (the highest bid listed in the book): so that
$p_t = (a^q_t + b^q_t)/2$, a value that we call the mid-point. If no
bids or asks are listed in the book a proxy for the price is given
by the previous traded or quoted price. Bids, asks and prices need
to be positive and investors can submit limit orders at any price on
a prespecified grid, defined by the tick size $\Delta$.

The demands of each trader for the risky asset are assumed to
consist of three components, a fundamentalist component, a chartist
component and a noise induced component. The weights applied to
these various components will vary in ways to be described below. At
any time $t$ a trader is chosen to enter the market. The chosen
agent, $i$, forms an expectation about the spot return, $\hat
r^i_{t,t + \tau^i}$, that will prevail in the interval
$(t,t+\tau^i)$, where $\tau^i$ is the agent's time horizon. Agents
use a combination of fundamental value and chartist rules to form
expectations on stock returns, so that
\begin{equation}
\hat{r}^i_{t, t + \tau^i} = \frac{1}{g_1^i + g_2^i + n^i}\left[g_1^i\frac{1}{\tau_f}\ln(p^f_t/p_t)+ g_2^i \bar r^i_{t} + n^i \epsilon_t\right],
\label{expect}
\end{equation}
where the quantities $g_1^i > 0$ and $g^i_2>0$ represent the weights given to the fundamentalist and chartist component respectively. In addition, we add a noise induced component $\epsilon_t$, with zero mean and variance $\sigma_{\epsilon}$ to agent's expectations with weight $n^i>0$. The initial term in Eq.~(\ref{expect}) normalises the impact of the three trading strategies. The quantity $\tau_f$ is the time scale over which the fundamentalist component for the mean reversion of the price to the fundamental is calculated. Finally, the average $\bar r^i_{t}$ gives the future expected trend of the chartist component based on the observations of the spot returns over last $\tau^i$ time steps. That is,
\begin{equation}
\bar r^i_{t} = \frac{1}{\tau^i} \sum_{j=1}^{\tau^i} r_{t-j} = \frac{1}{\tau^i} \sum_{j=1}^{\tau^i} \ln
\frac{ p_{t-j}}{ p_{t-j-1}},
\label{ave}
\end{equation}
where $r_{t}$ and $p_t=p_{t-1}\exp(r_t)$ are, respectively, the spot return and the spot price at time $t$.

Observe that a pure fundamentalist trading strategy has $g_2^i =n^i=0$, a chartist trading strategy has $g_1^i=n^i=0$, whilst for a noise trading strategy $g_1^i =g^i_2 =0$. We assume that the degree of fundamentalism, chartism and noise trading will be spread across agents, so we model the trading weights as random variables independently chosen. For agent $i$ the weights are chosen according to the realizations of the set of Laplace distributions,
\begin{eqnarray}
\mbox{Prob}(g_1^i)=\frac{1}{\sigma_{1}}\exp(-g_1^i/\sigma_{1}),\nonumber\\
\mbox{Prob}(g_2^i)=\frac{1}{\sigma_{2}}\exp(-g_2^i/\sigma_{2}),\nonumber
\end{eqnarray}
and
%\begin{equation}
$$
\mbox{Prob}(n^i)=\frac{1}{\sigma_{n}}\exp(-n^i/\sigma_{n}),
%\end{equation}
$$
for $g_1^i,g_2^i$ and $n^i\geq 0$. The average and variance of these
densities depend on the values of $\sigma_1$, $\sigma_2$ and
$\sigma_n$. So that: $\mathbb{E}[g_1^i]=\sigma_{1}$ with variance
$\sigma_1^2,\mathbb{E}[g_2^i]=\sigma_2$ with variance $\sigma_2^2$
and $\mathbb{E}[n^i]=\sigma_{n}$ with variance $\sigma_{n}^2$.

The future price, $p^i_{t+\tau^i}$, expected at time $t + \tau^i$ by agent $i$ is given by
\begin{equation}
\hat p^i_{t + \tau^i} = p_t \exp(\hat{r}^i_{t, t +\tau^i} \tau^i),
\label{pexpect}
\end{equation}
where we recall that $p_t$ is the current price at time $t$. It is
common to assume that the time horizon of an agent depends on its
characteristics. Fundamentalist strategies are typically given much
greater weight by long term institutional investors who have longer
time horizons, whilst day traders have shorter time horizons and
give more weight to chartist rules. Hence we choose the time horizon
$\tau^i$ of each agent according to\footnote{The notation $[x]$
refers to the value obtained by rounding up to the next highest
integer.}
\begin{equation}
\tau^i = \left[\tau \frac{1+g_1^i}{1+ g_2^i} \right],
\label{tau}
\end{equation}
where $\tau$ is some reference time horizon.

Once the agent has formed its own expectation of the future price,
it has to decide whether to place a buy or a sell order and choose
the size of the order. We assume that agents are risk averse and
maximize the exponential utility of wealth function\footnote{We
recall that this utility function belongs to the CARA (constant
absolute risk aversion) class.}
\begin{equation}
U(W^i_t,\alpha^i) = -e^{-\alpha^i W^i_t},
\label{utility}
\end{equation}
where the coefficient $\alpha^i$ measures the relative risk aversion
of agent $i$. We assume that those agents giving greater weight to
the fundamentalist component are more risk averse than those giving
more weight to the noise trader and chartist components. This effect
is captured by setting
\begin{equation}
\alpha^i = \alpha \frac{1+g_1^i}{1+g_2^i},
\label{alpha}
\end{equation}
where $\alpha$ is some reference level of risk aversion.

We next define the portfolio wealth of each agent as
\begin{equation}
W^i_t=S^i_t p_t+ C^i_t,
\label{W}
\end{equation}
where $S^i_t\geq 0$ and $C^i_t\geq 0$ are respectively the stock and
cash position of agent $i$ at time $t$. The optimal composition of
the agent's portfolio is determined in the usual way by trading-off
expected return against expected risk. However the agents are not
allowed to engage in short-selling. The number of stocks an agent is
willing to hold in its portfolio at a given price level $p$ depends
on the choice of the utility function. For the CARA utility function
assumed here the optimal composition of the portfolio, that is the
number of stocks the agent wishes to hold is given by
\begin{equation}
\pi^i(p)=\frac{\ln(\hat{p}^i_{t+\tau^i}/p)}{\alpha^i
V^i_t p},
\label{equation8}
\end{equation}
where $V_t^i$ is the variance of returns expected by agent $i$, the
agent's relative risk aversion $\alpha^i$ is given by
Eq.~(\ref{alpha}), and the agent's investment horizon $\tau_i$ is
defined in Eq.~(\ref{tau}). We note that Eq.~(\ref{equation8}) is
independent of wealth, which obviates the need to keep track of the
wealth dynamics of each individual agent. If the amount $\pi^i(p)$
is larger (smaller) than the number of stocks already in the
portfolio of agent $i$ then the agent decides to buy (sell).
Eq.~(\ref{equation8}) can be derived on the basis of mean-variance
one-period portfolio optimization. Appendix~\ref{demand} shows how
to obtain this equation from the utility function~(\ref{utility})
following a similar approach to that of Bottazzi et al. (2005). We
estimate $V^i_t$ as the variance of past returns, estimated by each
agent as
\begin{equation}
V^i_t = \frac{1}{\tau^i}\sum_{j=1}^{\tau^i} [r_{t-j} - \bar r^i_t]^2,
\label{sigma}
\end{equation}
where the average spot return $\bar r^i_t$ is given by Eq.~(\ref{ave}).

In order to determine the buy/sell price range of a typical agent,
we first estimate numerically the price level $p^*$ at which agents
are satisfied with the composition of their current portfolio, which
is determined by
\begin{equation}
\pi^i(p^*) = \frac{\ln(\hat p^i_{t+\tau^i}/p^*)}{\alpha^i V^i_t p^*} = S^i_t.
\label{p*}
\end{equation}
Eq.~(\ref{p*}) admits a unique solution with $0< p^*\leq\hat p^i_{t
+ \tau^i}$ since $S_t^i \geq 0$ given that short selling is not
allowed. Agents are willing to buy at any price $p<p^*$ since in
this price range their demand is greater than their holding. While
agents are willing to sell at any price $p>p^*$ since then their
demand is less than their holding. Note that agents may thus wish to
sell even if they expect a future price increase. If we select
$p=p^*$ agents decide to do nothing.

\begin{figure}[t]
\begin{center}
\includegraphics[width=6cm]{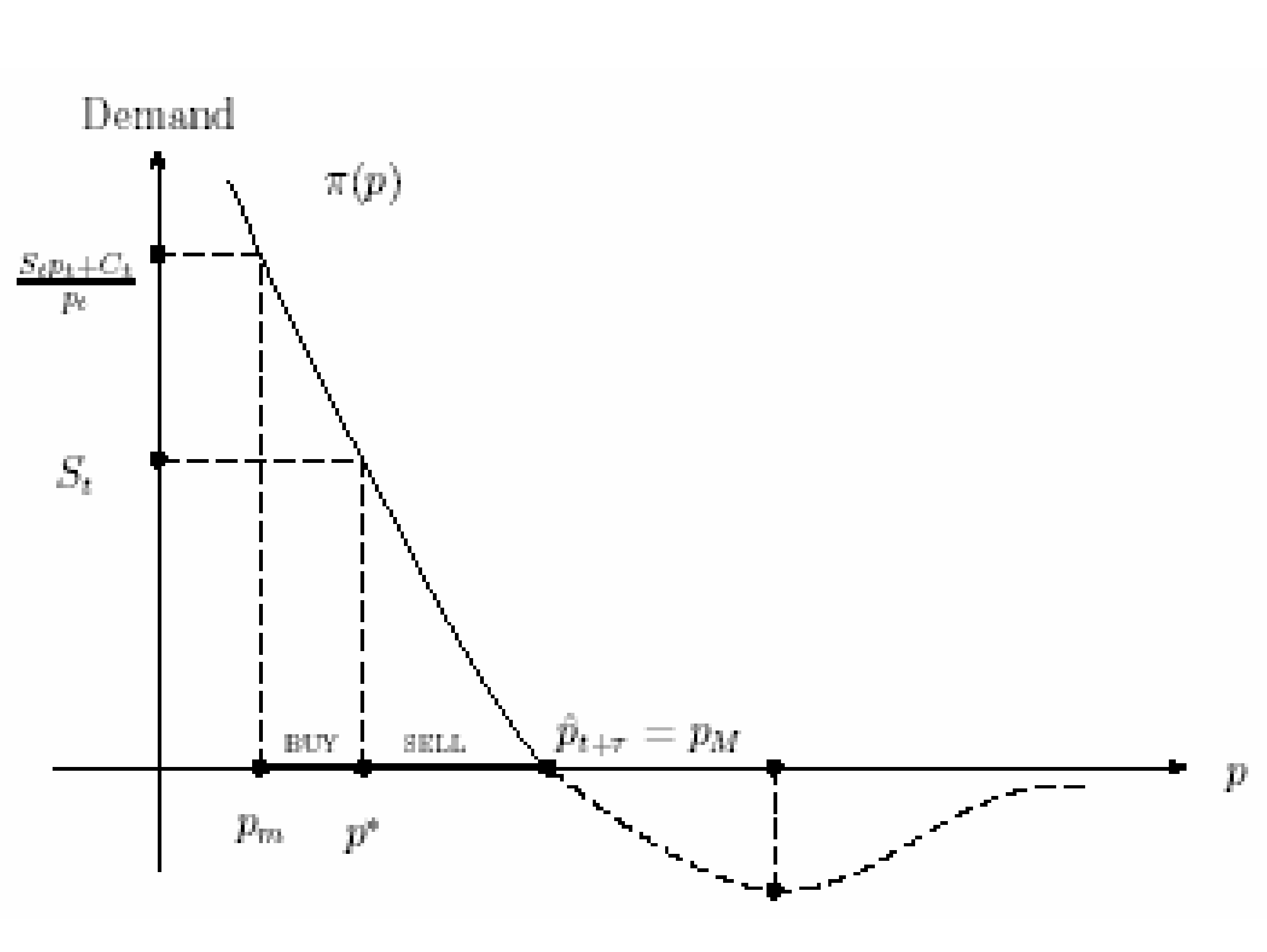}
\caption{Determination of the buy/sell price range of a typical agent (the superscript $i$ has been suppressed here). The curve labelled $\pi(p)$ graphs Eq.~(\ref{equation8}) with the dashed section representing negative (short position) demand. The upper limit of the price range is $p_M$ where the positive (long position) demand comes to zero. The minimum limit $p_m$ is determined by the agent's current wealth holding. The price $p^*$ that separates the buy/sell regions is determined by the agent's current holding, $S_t$, of stock.}
\label{figure}
\end{center}
\vspace*{13pt}
\end{figure}

As we want to impose budget constraints we need to restrict ourselves to values of $p \leq \hat p^i_{t + \tau^i}=p_M$ to ensure $\pi(p)\geq 0$ and so rule out short selling. Furthermore to ensure that an agent has sufficient cash to purchase the desired stocks, the smallest value of $p=p_m$ we can allow for agent $i$ is
determined by its cash position (see Eq.~(\ref{W})), and so is given by the condition
\begin{equation}
p_m\left(\pi^i(p_m) - S^i_t\right) = C^i_t.
\label{pm}
\end{equation}
Again one can easily show that this equation also admits a unique
solution with $0<p_m\leq\hat p^i_{t + \tau^i}$ since
$S^i_t,C^i_t\geq 0$. Indeed, comparing Eqs.~(\ref{p*})
and~(\ref{pm}) it can be easily proven that $0<p_m\leq p^*\leq \hat
p^i_{t+\tau^i}$. We represent the typical agent's buy/sell price
range graphically in Fig.~\ref{figure}.

\begin{table}[tbp]
\begin{center}
\begin{tabular}{lllll}
\hline
\hline
\\
 & Position & Type of order & Volume \\
\hline
\\
$p_m<p<a_t^q$ & BUY & Limit order & $s_i=\pi^i(p)-S^i_t$\\
$a_t^q\leq p<p^* $ & BUY & Market order & $s_i=\pi^i(a_t^q)-S^i_t$ \\
$p=p^*$ & No order placement \\
$p^*< p \leq b_t^q$ & SELL & Market order & $s_i=S^i_t-\pi^i(b_t^q)$\\
$b_t^q< p \leq p_M$ & SELL & Limit order & $s_i=S^i_t-\pi^i(p)$\\
\\
\hline \hline
\end{tabular}
\end{center}
\caption{Summary of the trading mechanism of a typical trader $i$
with a random price level $p$ limited between the value $p_m$ given
by Eq.~(\ref{pm}) and the value $p_M=\hat{p}_{t+\tau^i}^i$. The
current quoted best ask and best bid are $a_t^q$ and $b_t^q$
respectively.} \label{mechanism} \vspace*{13pt}
\end{table}

Having determined that the possible values at which an agent can
satisfactorily trade are in the interval $[p_m, p_M],$ we next
consider how the nature of the agent's order is determined. Agents
randomly draw a price $p$ from the interval $[p_m, p_M]$ and if
$p<p^*$ they submit a limit order to buy an amount
$$
s^i= \pi^i(p)-S^i_t,
$$
while if $p>p^*$ they submit a limit order to sell an
amount
$$
s^i= S^i-\pi^i(p).
$$
However if $p< p^*$ and $p> a^q_t$ the buy order can be executed immediately at the ask. An agent in this case would submit a market order to buy an amount
$$
s^i= \pi^i(a^q_t)-S^i_t.
$$
Similarly if $p > p^*$ and $p < b^q_t$ the agent would submit a market order to sell an amount
$$
s^i= {S^i_t}-\pi^i(b^q_t).
$$
If the depth at the bid (ask) is not enough to fully satisfy the
order, the remaining volume is executed against limit orders in the
book. The agent thus takes the next best buy (sell) order and
repeats this operation as many times as necessary until the order is
fully executed. This mechanism applies under the condition that
quotes of these orders are above (below) price $p$. Otherwise, the
remaining volume is converted into a limit order at price $p$. If
the limit order is still unmatched at time $t+\tau^i$ it is removed
from the book.

The essential details of the trading mechanism are summarized in Table~\ref{mechanism}, showing how it depends on the price level $p$, the ``satisfaction level'' $p^*$, the best ask $a_t^q$ and the
best bid $b_t^q$.

\section{Simulation Analysis}\label{sim anal}

In this section we analyze, via numerical simulations, various
properties of prices, order flows and the book implied by the model
of section 2. In the simulations we have considered (in succession)
three kinds of trading rules, noise trading only (black),
fundamentalism and noise trading only (red), and, fundamentalism,
chartism and noise trading (green). We set the number of agents at
$N_A=5000$. Agents are initially assigned a random amount of stock
uniformly distributed on the interval $S_0\in [0, N_S]$ and an
amount of cash $C_0 \in [0, W]$. In the following simulations we
choose $N_S=50$ and $W=N_S ~p_f(0)$. We also fix $\tau=200$,
$\Delta=0.0005$, $\sigma_{\epsilon}=10^{-4}$, $\sigma_n=1$, and
$\alpha$ = 0.1. We choose the fundamental value to be a random walk
with initial value $p_f(0)=300,$ zero drift and and volatility
$\sigma_f = 10^{-3}$. We run the simulations with a large set of
values $\sigma_1$ and $\sigma_2$ ranging from 0 to 30 in order to
study the impact of different fundamentalist and chartist components
on price, order flows and the book.

The results reported here are the outcome of simulations of 200,000 steps, each of which we repeat 100 times with different random seeds. To test the robustness of the results we have repeated the simulations varying the parameter set within a small neighborhood of those used here. We have found that the qualitative features
reported below are fairly robust to such variations.

\subsection{Price and Returns}

\begin{figure}[t]
\vskip 2cm
\includegraphics[width=5cm]{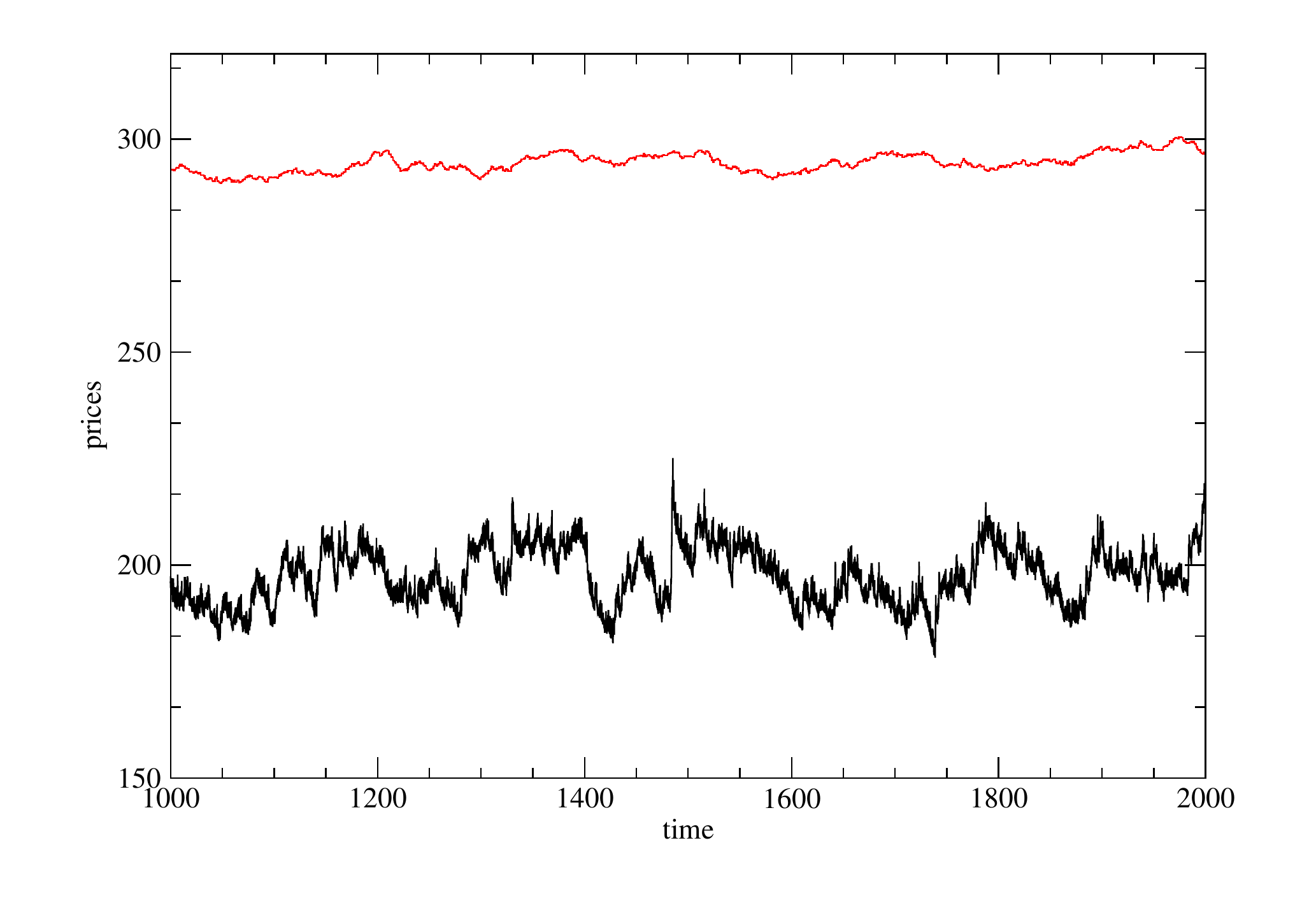}
\includegraphics[width=5cm]{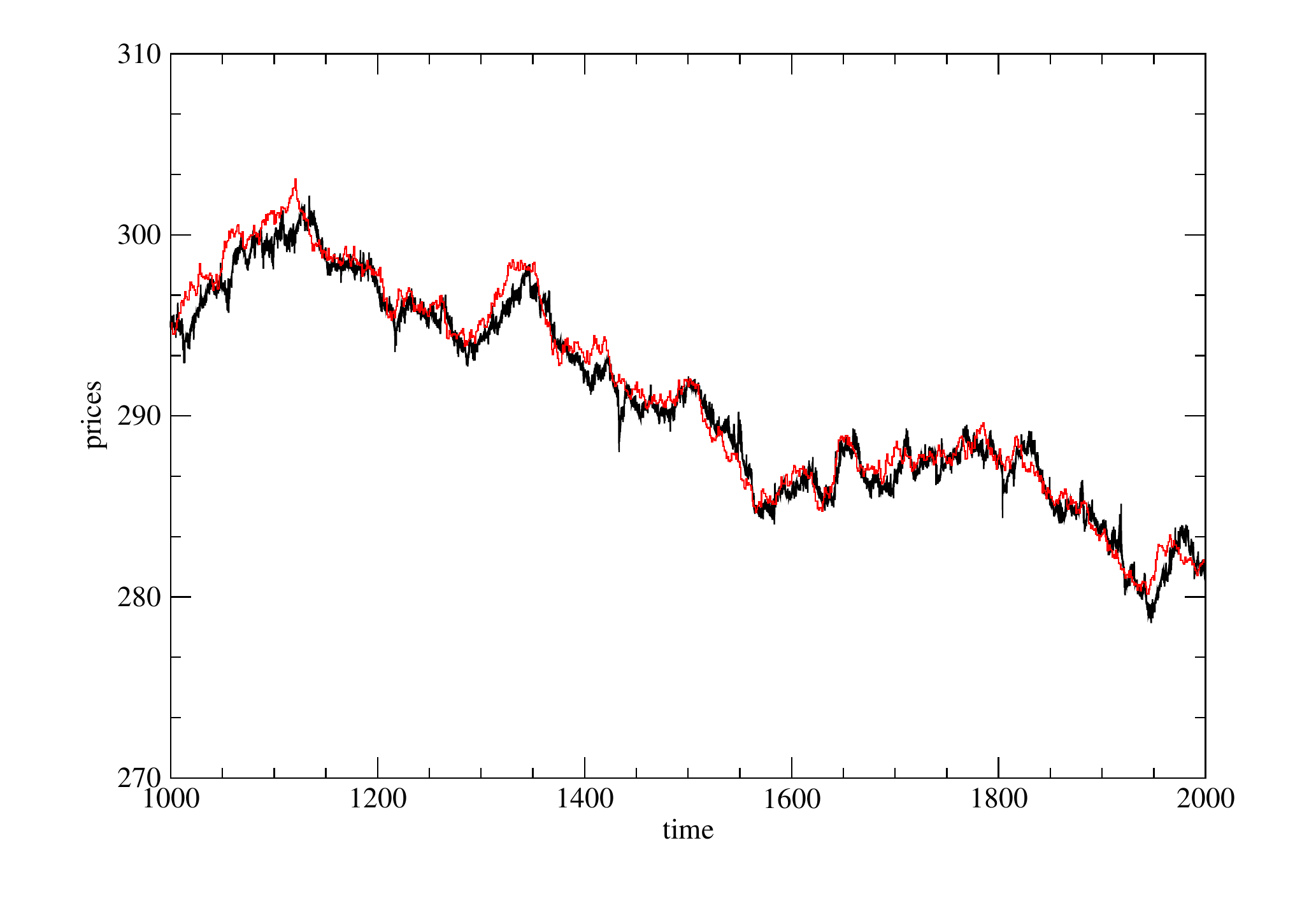} \includegraphics[width=5cm]{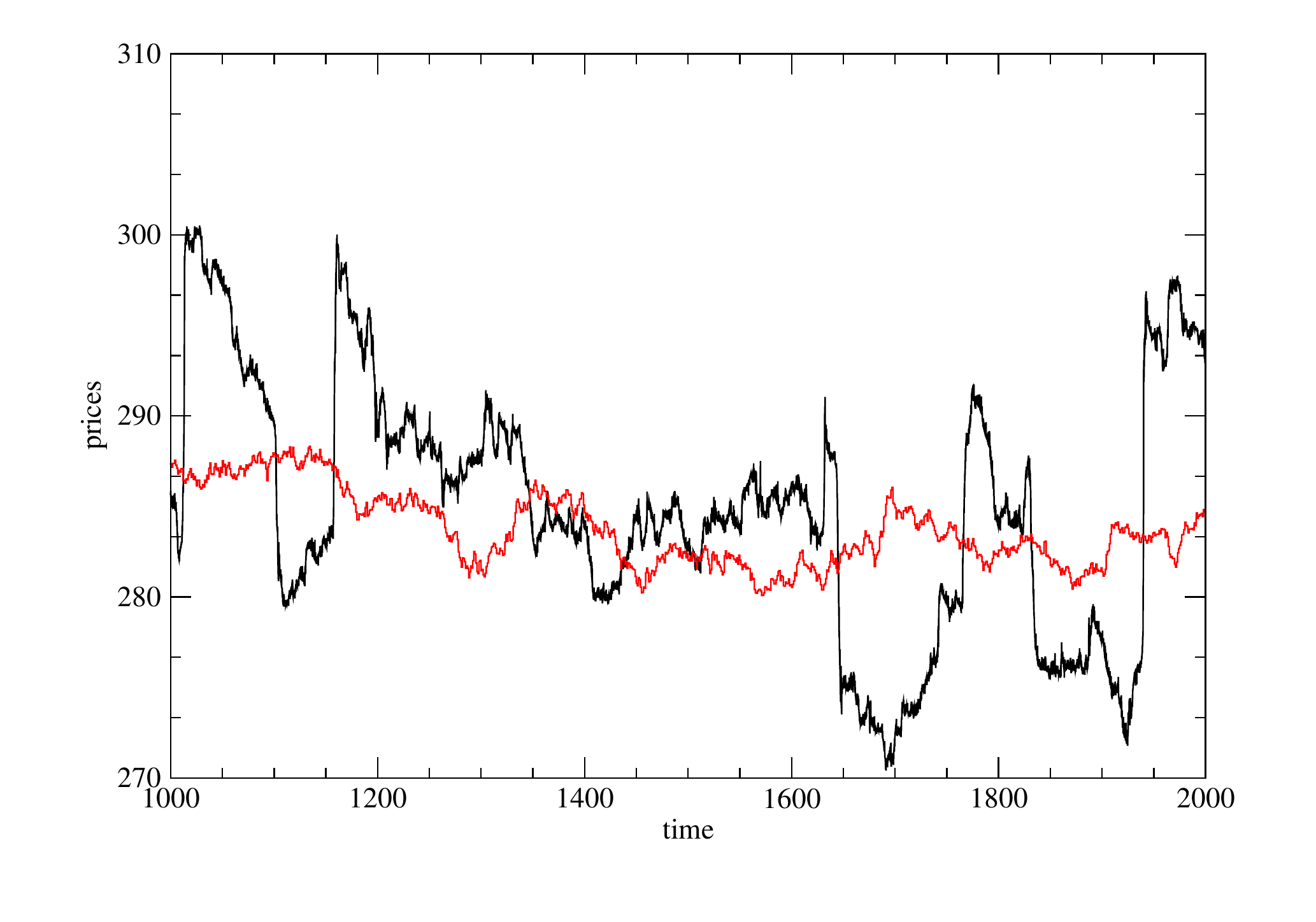}
\caption{Price (black) and fundamental value (red) with only noise trading (left, $\sigma_1=0.00$, $\sigma_2=0.00$), noise trading and fundamentalism (center, $\sigma_1 = 10.00$, $\sigma_2 = 0.00$) and noise trading, fundamentalism and chartism (right, $\sigma_1 = 10.00$, $\sigma_2 = 1.20$).}
\label{figure2}
\end{figure}

Our main aim in this section is to gain some insights into the
details of the price formation in the model. Figure \ref{figure2}
displays sample paths of the price under very different assumptions
concerning the trading strategies used. In the first path (left hand
side) the trading strategy only has a noise trading component and,
as we might expect, the fundamental price has nothing to do with the
current price generated by the double auction market. However, once
we include the fundamentalist component, the auction market is no
longer operating independently of the fundamental and we observe in
the centre plot that the market arrives at a price that follows
closely the fundamental price $p_f$. Finally, the right hand plot
shows how the addition of the chartist component to the trading
strategy affects the price evolution. This component accentuates the
fluctuations of the price as evidenced by the numerous extreme
movements that are indeed characteristic of real markets. These
simulations would suggest that big jumps in the price are mainly
caused by the chartist component.

\begin{figure}[t]
\includegraphics[width=6cm]{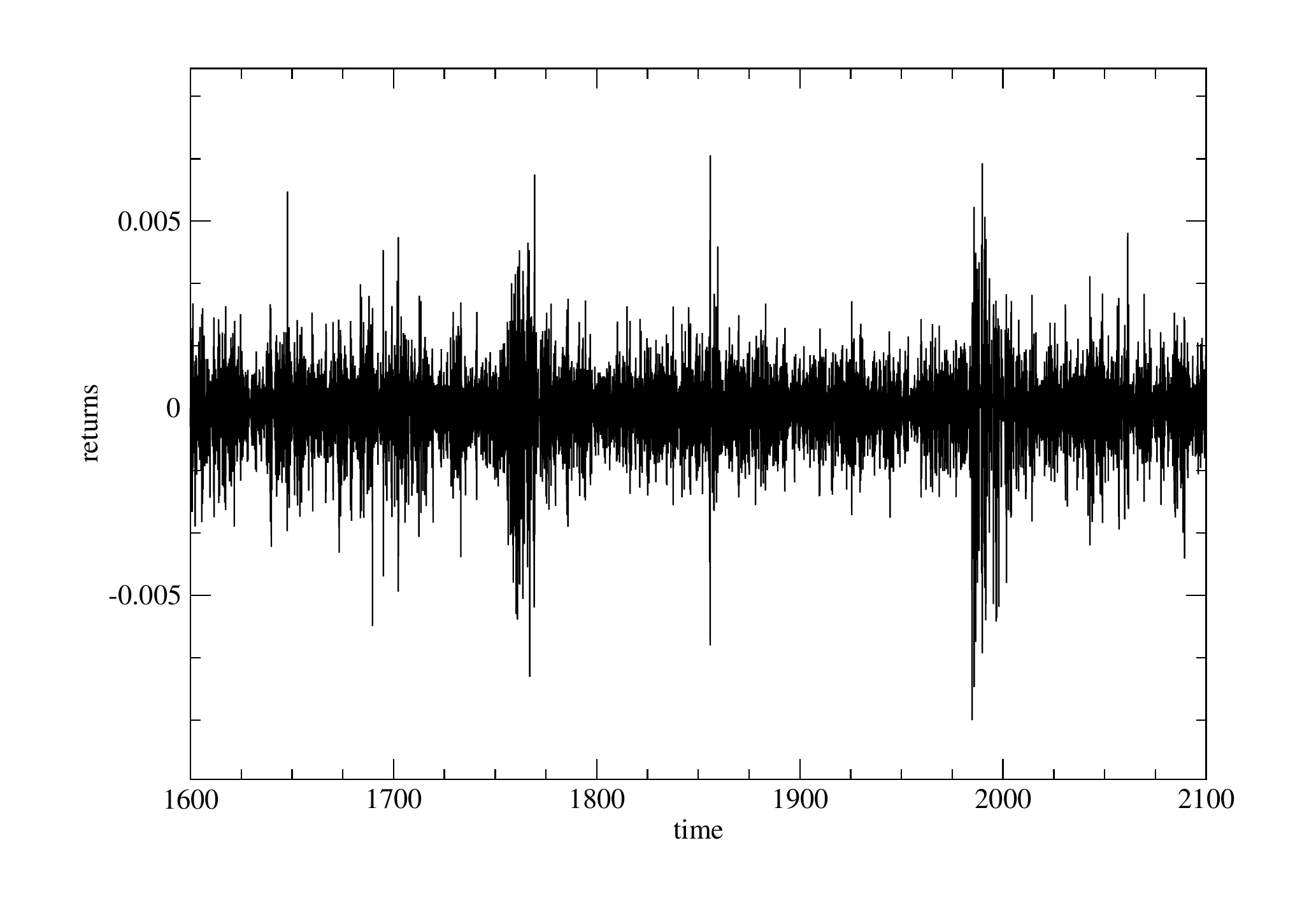}
\includegraphics[width=6cm]{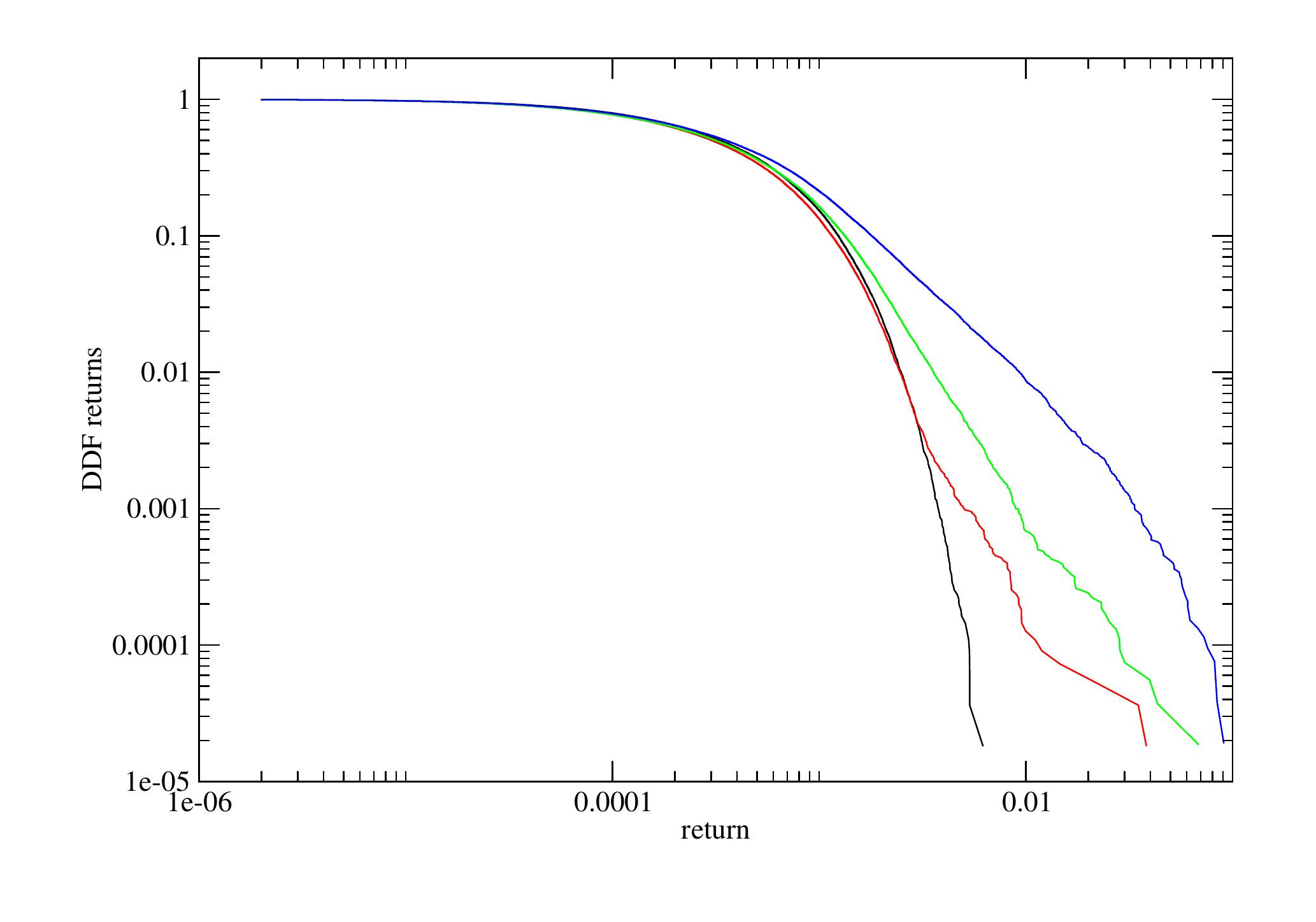}
\caption{(Left) Return time series with noise trader, fundamentalist and chartist ($\sigma_1 = 10.00$, $\sigma_2 = 1.20$) components all present. (Right) The decumulative distribution function of the absolute returns with $\sigma_1 = 10.00$ and $\sigma_2=0.00$ (black), $\sigma_2=1.20$ (red), $\sigma_2=2.00$ (green), $\sigma_2=2.80$ (blue).}
\label{figure3}
\end{figure}

Next we analyse the return time series. Figure~\ref{figure3} shows
on the left a rather typical path for the returns generated by the
model when noise trader, fundamentalist and chartist components are
all present. On the right side of Figure~\ref{figure3} we represent
the decumulative distribution functions (DDF) defined as the
probability of having a change in price larger than a certain return
threshold. In other words, we are plotting one minus the cumulative
distribution function. In this way we observe that fat tail
behaviour appears once chartism is introduced into the trading
strategy and increases with $\sigma_2$. The four curves correspond
to $\sigma_1=10.00$ and $\sigma_2=0.00$ (black), $\sigma_2=1.20$
(red), $\sigma_2=2.00$ (green), $\sigma_2=2.80$ (blue).

\begin{figure}[t]
\includegraphics[width=6cm]{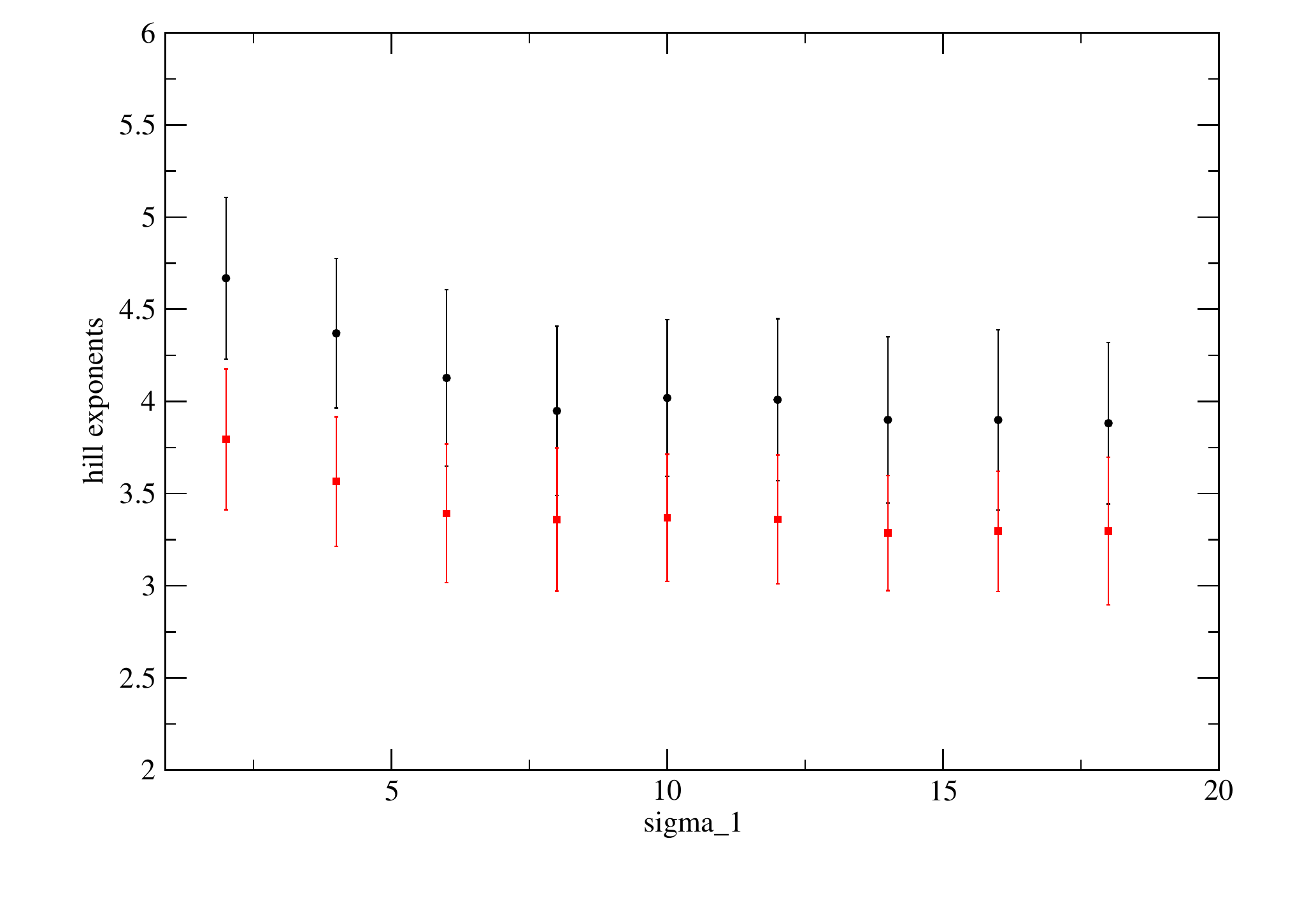}
\includegraphics[width=6cm]{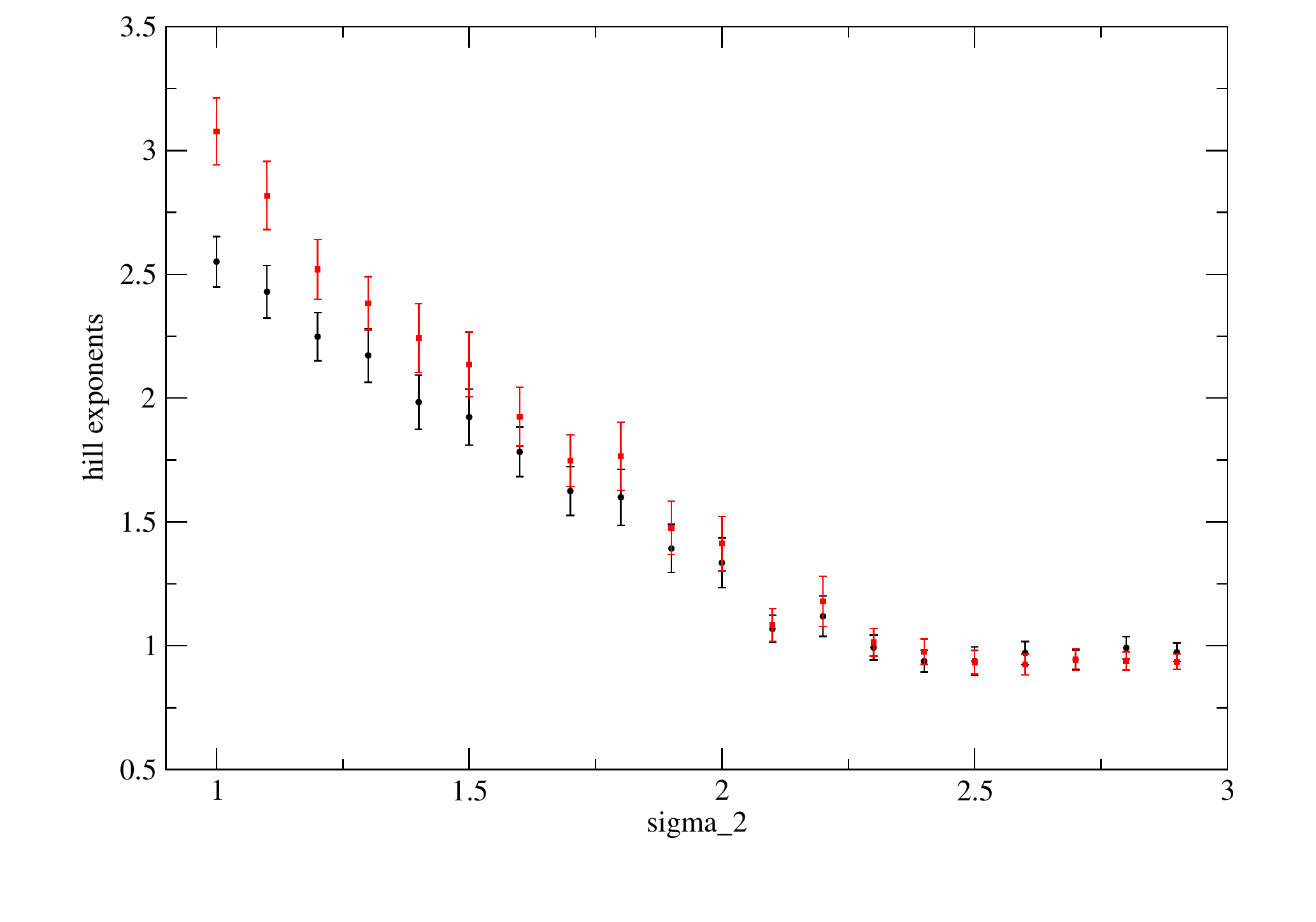}
\caption{Hill exponents of the left (black) and right (red) tails of the returns distribution as a function of $\sigma_1$ with $\sigma_2=0.00$ (left) and as a function of $\sigma_2$ with $\sigma_1=10.00$ (right). Vertical spreads depicts the error bars for the Hill exponent which are evaluated across 100 repetitions of the simulations with different random seeds.}
\label{figure4}
\end{figure}

To confirm and quantify the robustness of all the above statements
we have also computed the Hill tail index, the details of which are
explained in Appendix~\ref{Hill}. Essentially this index measures
the tail exponent (Hill (1975)), the lower it is the fatter the tail
of the DDF. The results are plotted in Figure~\ref{figure4}. The
left hand plot shows how the inclusion of the fundamentalist
component brings about a fat tail distribution although we do not
see big changes when we increase the fundamentalist weight
$\sigma_1$ in the market. The index decreases very slowly and takes
values between 4.5 and 3.5, while in real markets the tail index is
normally around or even below 3 (Lux (2001), Gabaix et al. (2003)).
We also detect that the right tail (due to positive changes in
returns) of the distribution is fatter (even though the differences
are not statistically significant enough) for all values of
$\sigma_1$. This would be in contradiction with the situation in
real markets where the skewness of the returns is found to be
negative indicating that the tail due to negative returns is fatter
than that due to positive returns (Lux (2001)).

We consider the impact of the chartist component on the right hand
side plot of Figure~\ref{figure4}. We observe a much clearer pattern
when the weight of this component of trading behavior increases. The
tail index diminishes drastically to a value below 3 as we increase
the weight of the chartist component $\sigma_2$. The second
important point to emerge from this plot is that it provides the
consistent negative skewness for the pdf observed in real markets.
The left tail of the distribution (negative returns) is fatter than
the right one. However, increasing the value of $\sigma_2$ beyond
about the value of 2 leads to a reversal of the skewness property in
the pdf, even though the tails become fatter. Clearly, the
parameters of the model need to have a significant chartist
component in order that it exhibit behaviour consistent with
empirical observations, but not so big as to lose the negative
skewness between the positive and negative tails of the return pdf.

\begin{figure}[t]
\vskip .5cm
\includegraphics[width=6cm]{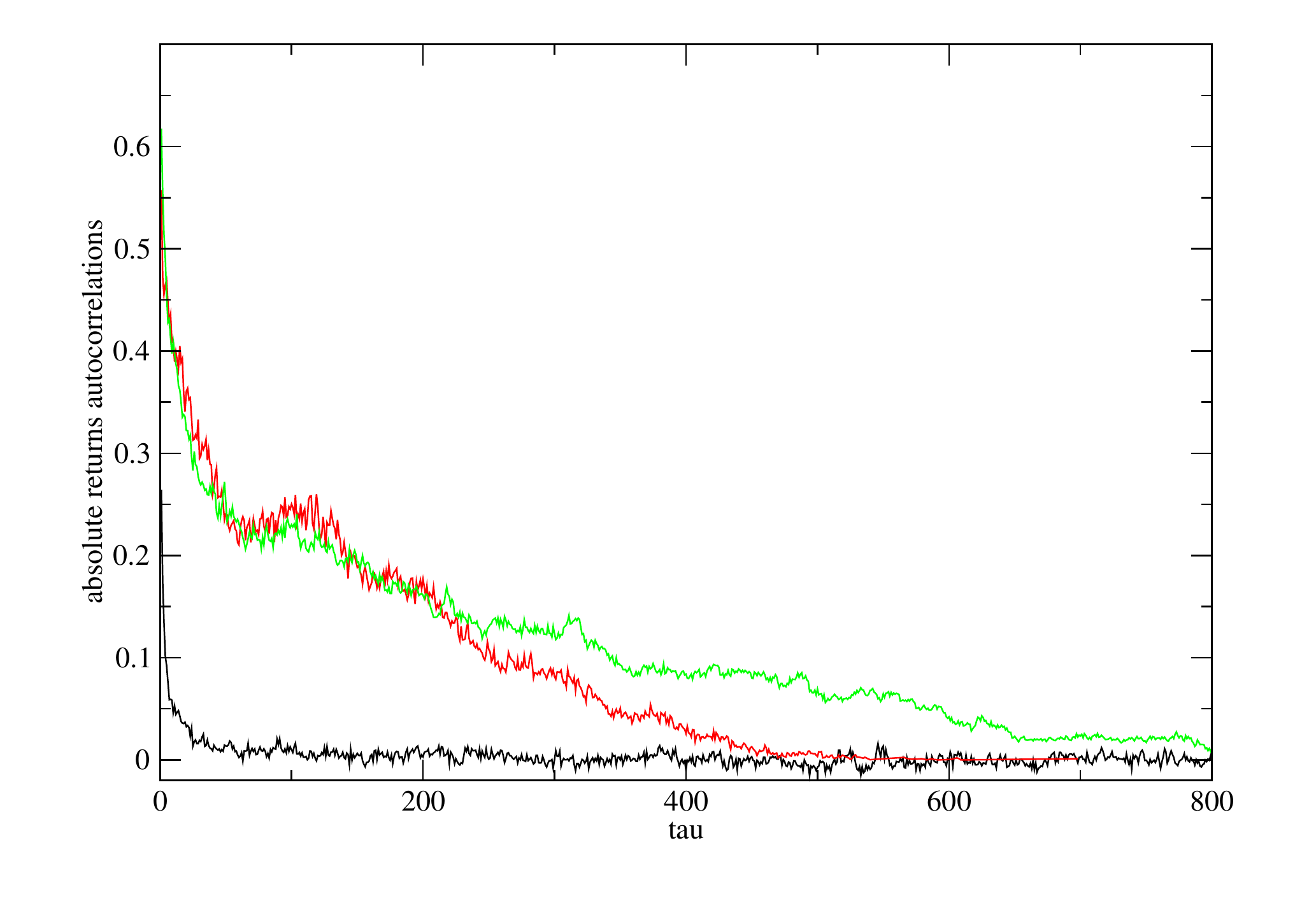}
\includegraphics[width=6cm]{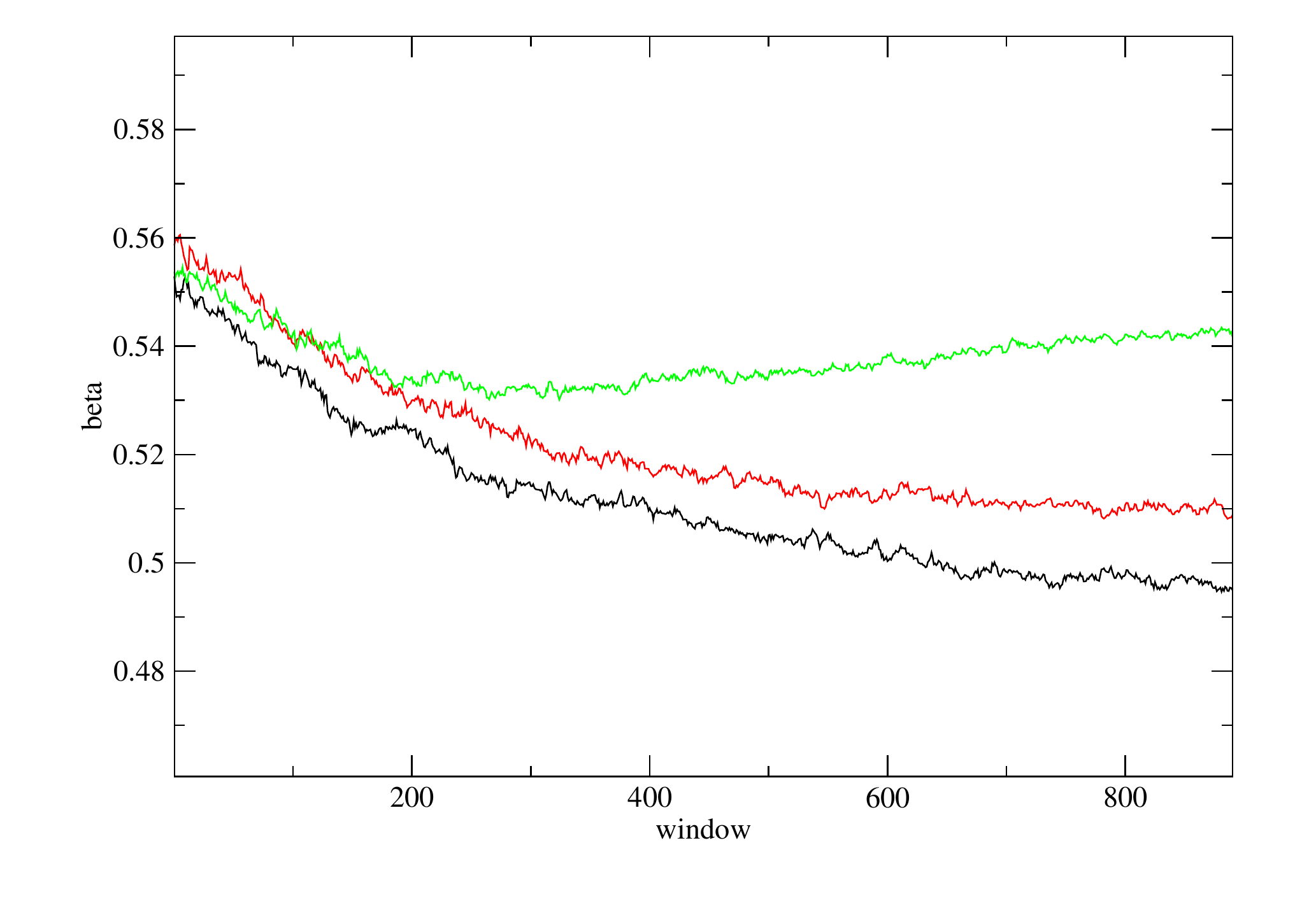}
\caption{(Left) Volatility autocorrelation decays for noise trading (black, $\sigma_1= \sigma_2=0.00$), noise
trading and fundamentalism (red, $\sigma_1=10.00$, $\sigma_2=0.00$) noise trading fundamentalism and chartism (green, $\sigma_1=10.00$, $\sigma_2=1.20$). (Right) The modified R/S exponent for absolute returns on the left side in terms of the time window. We here represent the exponent $\beta_n$ for several trading weights and filter the short term memory for a number of time steps $q=20$ (cf. Eq.~(\ref{RS1})).}
\label{figure5}
\end{figure}

We now focus on the long memory property of the volatility. We first
observe this property by means of the volatility autocorrelation
\begin{equation}
C_j=\frac{\hat{\gamma}_{j,N}}{\hat{V}_N}=\frac{\frac{1}{N}\sum_{i=j+1}^{N}\left(X_i-\bar{X}_N\right)\left(X_{i-j}-\bar{X}_N\right)}{\frac{1}{N}\sum_{j=1}^{N}\left(X_j-\bar{X}_N\right)^2},
\label{auto}
\end{equation}
where the $X_1,..., X_n,..., X_N$ are the absolute returns $|r_t|=|\ln(p_t/p_{t-1})|$ over time steps $1$ to $N$. The variable $\bar{X}_N$ is the absolute return sample mean over a time window of $N$ time steps
\begin{equation}
\bar{X}_N=\frac{1}{N}\sum_{i=1}^N X_i.
\label{mean}
\end{equation}
The autocorrelation thus calculates the ratio between the autocovariance and variance estimators. In Figure~\ref{figure5} (left) we plot the volatility autocorrelation for noise trading (black), noise trading
plus fundamentalism (red) and noise trading plus fundamentalism and chartism (green) in terms of $j$ and averaging over the whole data set generated from the simulations. The volatility autocorrelation is non-negligible when we include the fundamentalist and chartist components in the market for a broader domain. Real markets show a very long range volatility autocorrelation (Lo (1991)). Nonetheless, our model only shows very long term memory when the chartist component is activated. Hence, we not only observe a direct relationship between the chartist component and large returns but also detect that long memory in the volatility appears to be directly related to the inclusion of the chartism component.

To better quantify this effect we have studied the long range memory of the volatility by implementing the modified R/S, or rescaled range, analysis (see Figure~\ref{figure5}, right). Basically, the R/S statistic, introduced by Mandelbrot (1972) is the range of partial sums of deviations of a time series from its
mean, rescaled by its standard deviation. Lo (1991) has modified the statistic so that its behaviour is invariant for short memory processes but deviates for long memory process. The modified R/S statistic is defined as
\begin{equation}
Q_n=\frac{1}{\sqrt{\hat{V}_n(q)}}\left[\max_{1\leq k\leq n}\sum_{i=1}^{k}\left(X_i-\bar{X}_n\right)-\min_{1\leq k\leq n}\sum_{i=1}^{k}\left(X_i-\bar{X}_n\right)\right],
\label{RS1}
\end{equation}
where $X_1,..., X_n,..., X_N$, are a sample of absolute returns and
$$
%\begin{equation}
\hat{V}_n(q)=\hat{V}_n+2\sum_{j=1}^{q}\omega_j(q)\hat{\gamma}_{j,n},
%\label{RS2}
%\end{equation}
$$
with
$$
%\begin{equation}
\omega_j(q)=1-\frac{j}{q+1}, \qquad q<n.
%\label{omega}
%\end{equation}
$$
The term $\bar{X}_n$ appearing in Eqs.~(\ref{RS1}) and given by Eq.~(\ref{mean}) is the sample mean over the time window $n$. Also note that $\hat{V}_n$ and $\hat{\gamma}_{j,n}$ are the usual sample of variance and autocovariance estimators of X provided in Eq.~(\ref{auto}) but taking only $n$ elements of the whole sample of
size $N$. The original R/S estimator is recovered by setting $q=0$. With a value of $q>0$ the modified Lo estimator allows us to remove the effects of short range correlations up to the level $q$ and thus focuses on long memory effects in the data. The relevant quantity to study is the ratio
\begin{equation}
\beta_n=\frac{\ln Q_n}{\ln n},
\end{equation}
where $Q_n$ is given by Eq.~(\ref{RS1}). The value $\beta_n \neq
1/2$ indicates that there is still memory up to the time scale $n$.
If $\beta_n>1/2$ the series has a persistent behaviour (at least up
to the time scale $n$), thus being positively correlated. Otherwise,
if $\beta_n < 1/2$, the series has an anti persistent behaviour,
thus being negatively correlated.

The graphs of $\beta_n$ in the right hand panel of Fig.~\ref{figure5},
with $\beta_n > 1/2$ for window lengths up to 800 timesteps, again confirm that the
chartist component is an important factor in the long memory
property of return series.

\subsection{Book and Order flows}

In this section we investigate the impact of fundamentalist and
chartist trading profiles on the order submission strategies and the
resulting market book shape.

\begin{figure}[t]
\vskip 2cm
\includegraphics[width=6cm]{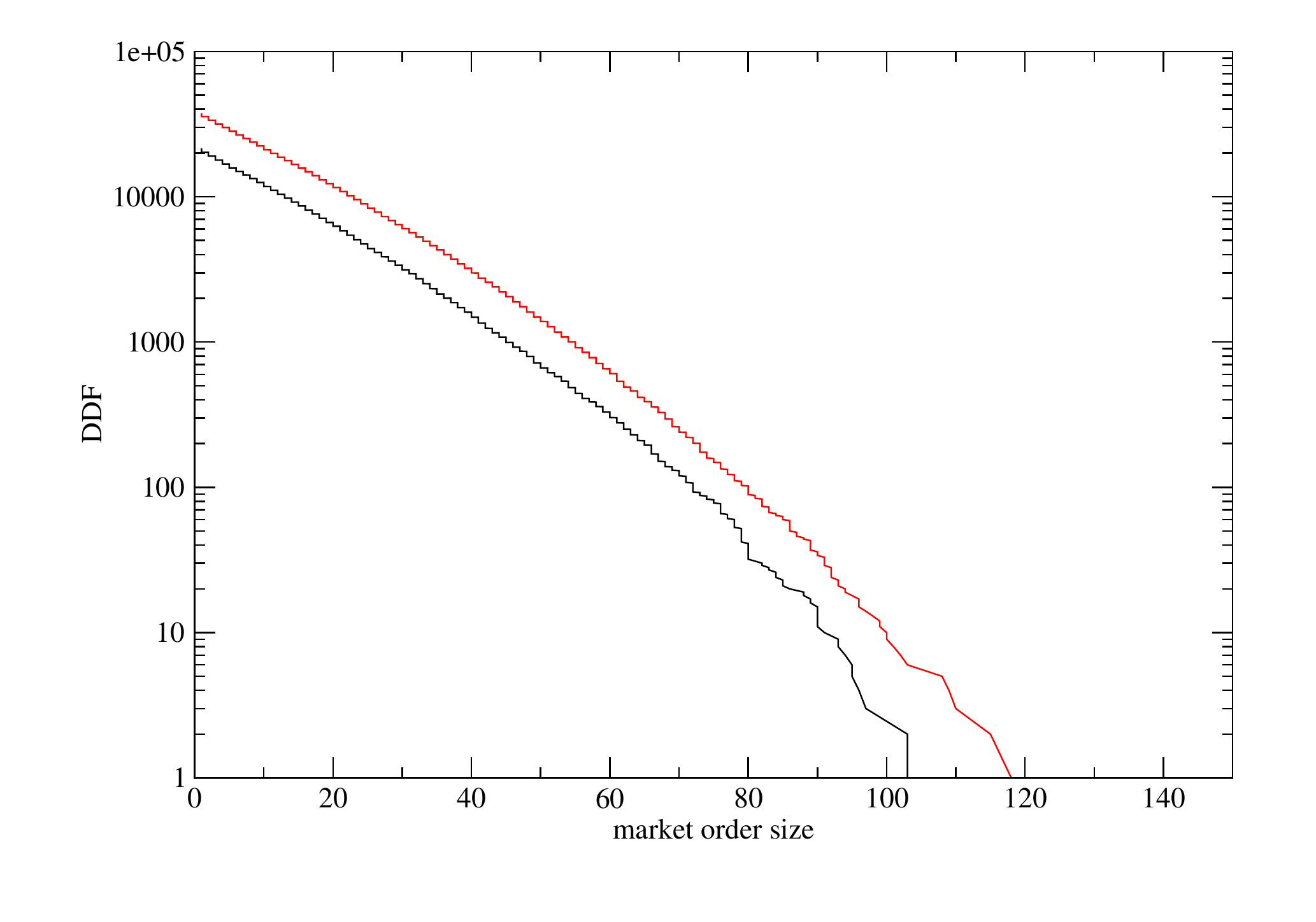}\includegraphics[width=6cm]{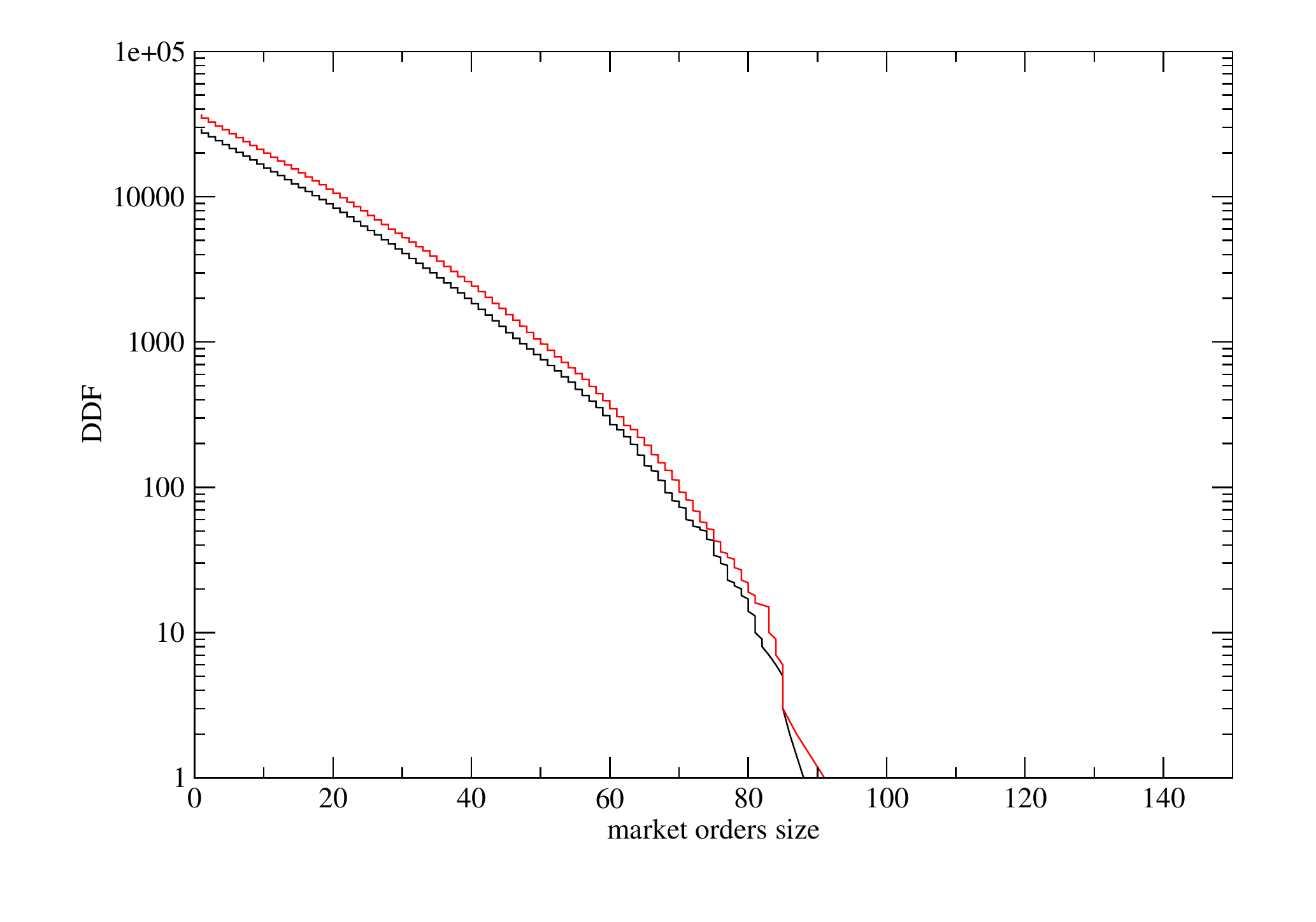}\includegraphics[width=6cm]{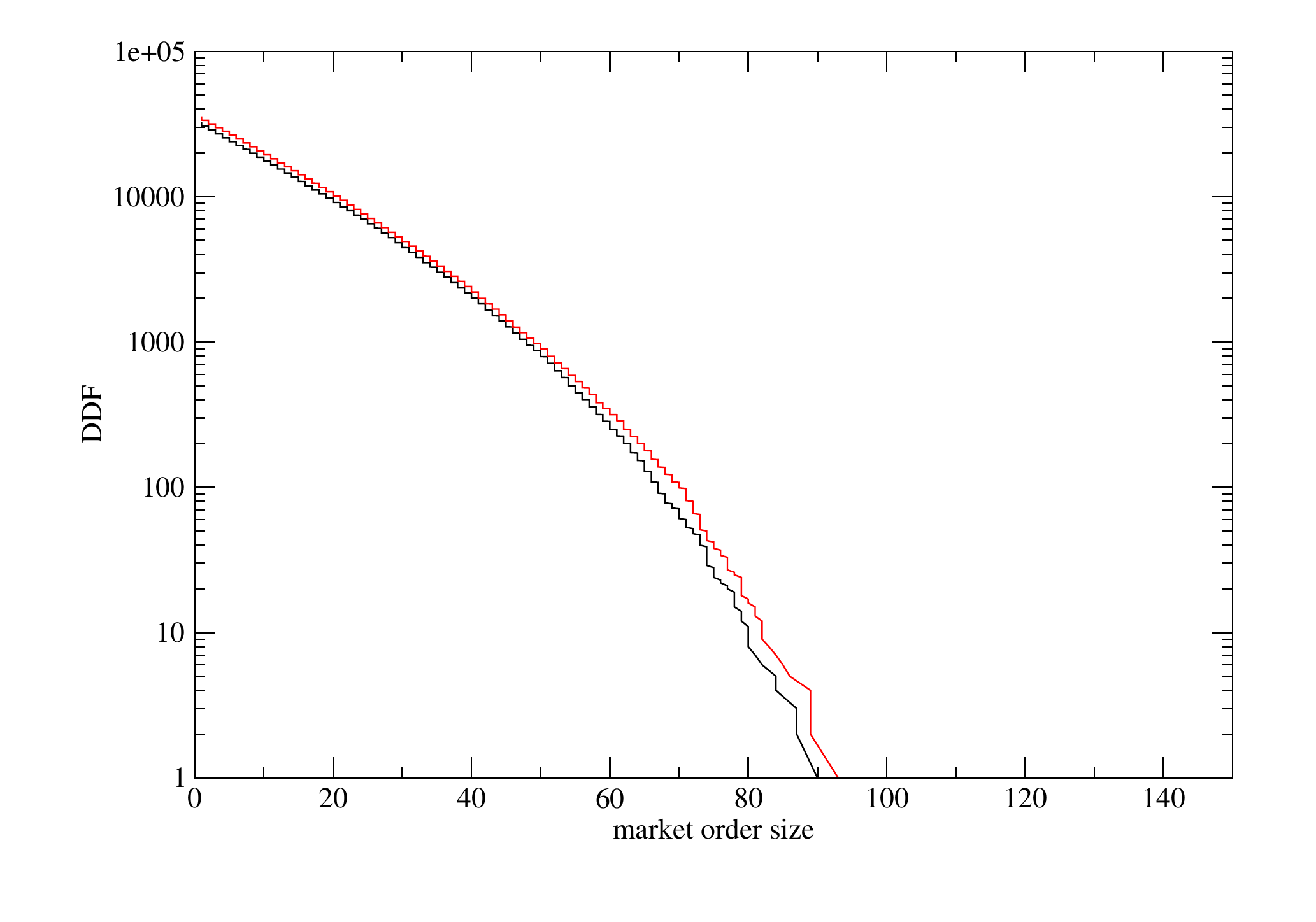}
\caption{The decumulative distribution function (not normalized) of market order size for buy (black) and sell (red) orders with noise trading (left), noise trading and fundamentalism (centre), and noise trading, fundamentalism and chartism (right). The value of the parameters are the same as in Figure \ref{figure5}.}
\label{figure6b}
\end{figure}

Figure~\ref{figure6b} reveals  a higher number, and larger size, of
market orders to sell with respect to market orders to buy,
particularly in the case of noise trader strategy only. This
indicates that agents  are more likely to demand  immediacy in
execution, via the use  of market orders, on the sell side of the
market.

\begin{figure}[t]
\vskip 2cm
\includegraphics[width=6cm]{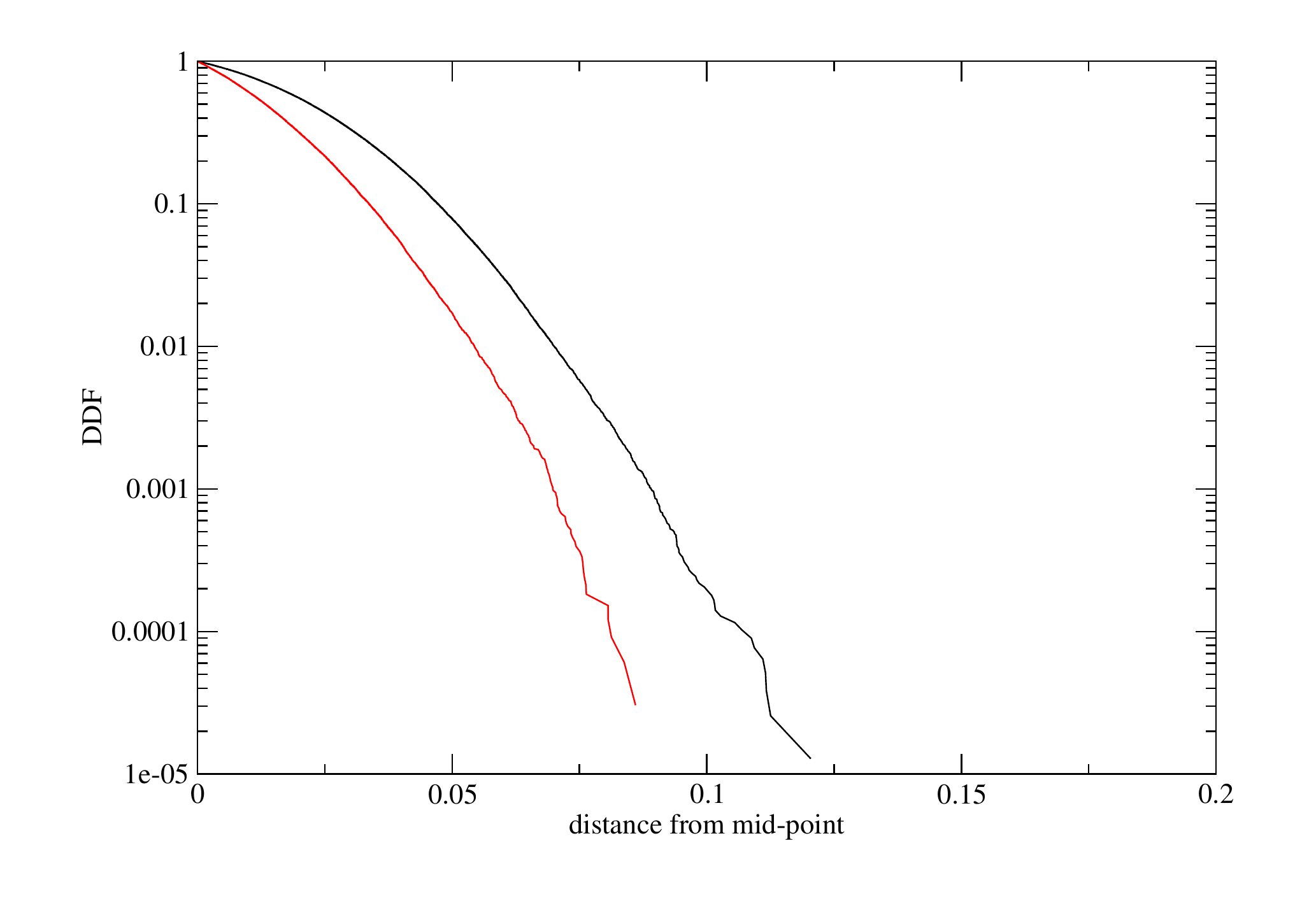}\includegraphics[width=6cm]{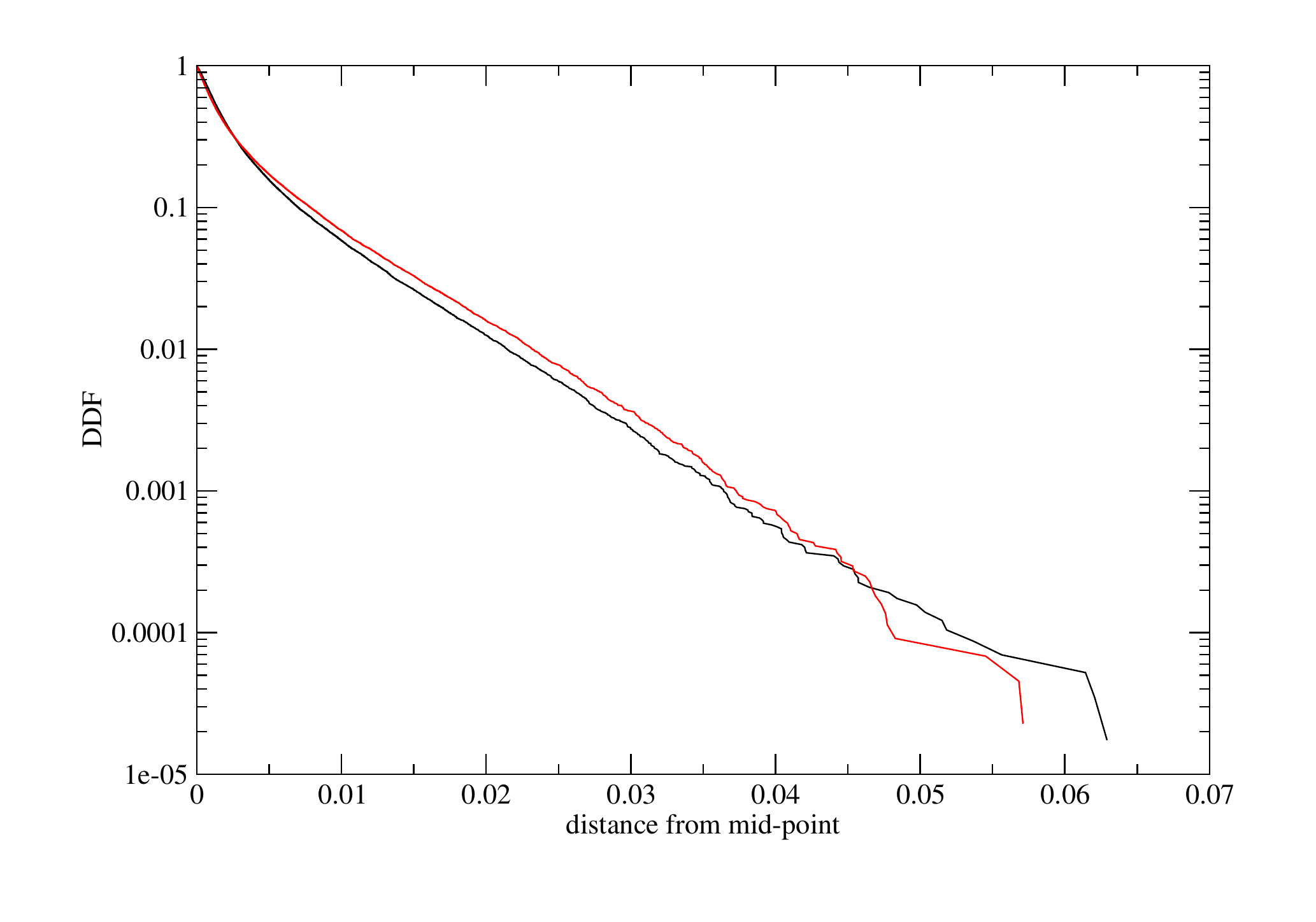}\includegraphics[width=6cm]{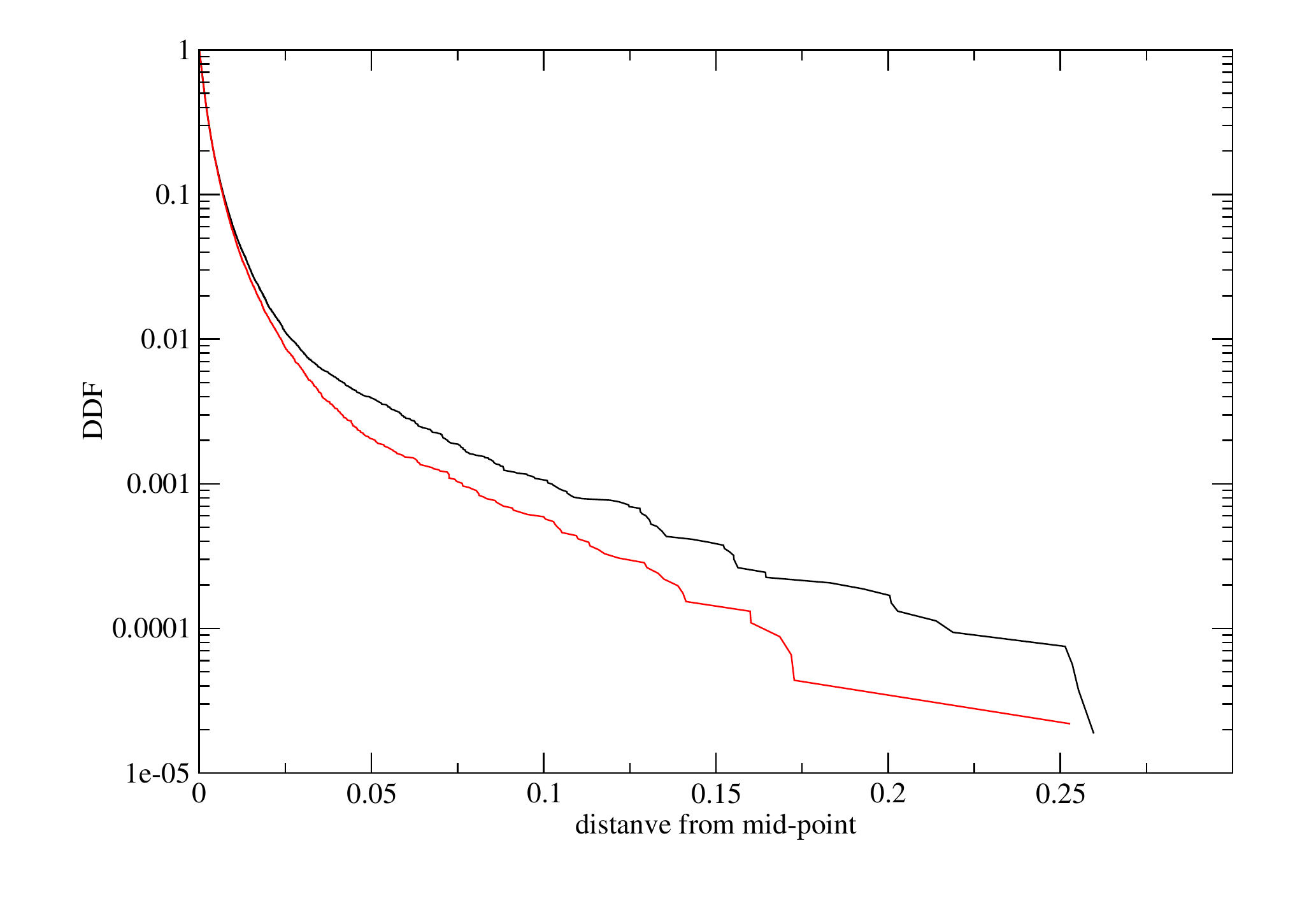}
\caption{The decumulative distribution function of limit order placement distance from the midpoint for buy (black) and sell (red) orders with noise trading (left), noise trading and fundamentalism (centre), and noise trading, fundamentalism and chartism (right). The value of the parameters are the same as in Figure~\ref{figure5}.}
\label{figure6}
\end{figure}

Figure~\ref{figure6} displays the decumulative distribution function (DDF) of the limit order placement distance from the midpoint, for buy (black) and sell (red) limit orders, with noise trading (left), noise trading and fundamentalism (centre), and noise trading, fundamentalism and chartism (right). We observe that with only noise trading, the DDF appears to be normally distributed for both buy and
sell orders. When the fundamentalist component dominates both buy
and sell limit orders are placed very close to the midpoint and the
DDF appears to have an exponential decay (this also explains why the
price follows very closely the fundamental price in this case). When
the chartist component is activated more and more limit orders are
placed at larger distances from the midpoint generating a hyperbolic
decay for the DDF.

\begin{figure}[t]
\includegraphics[width=6cm]{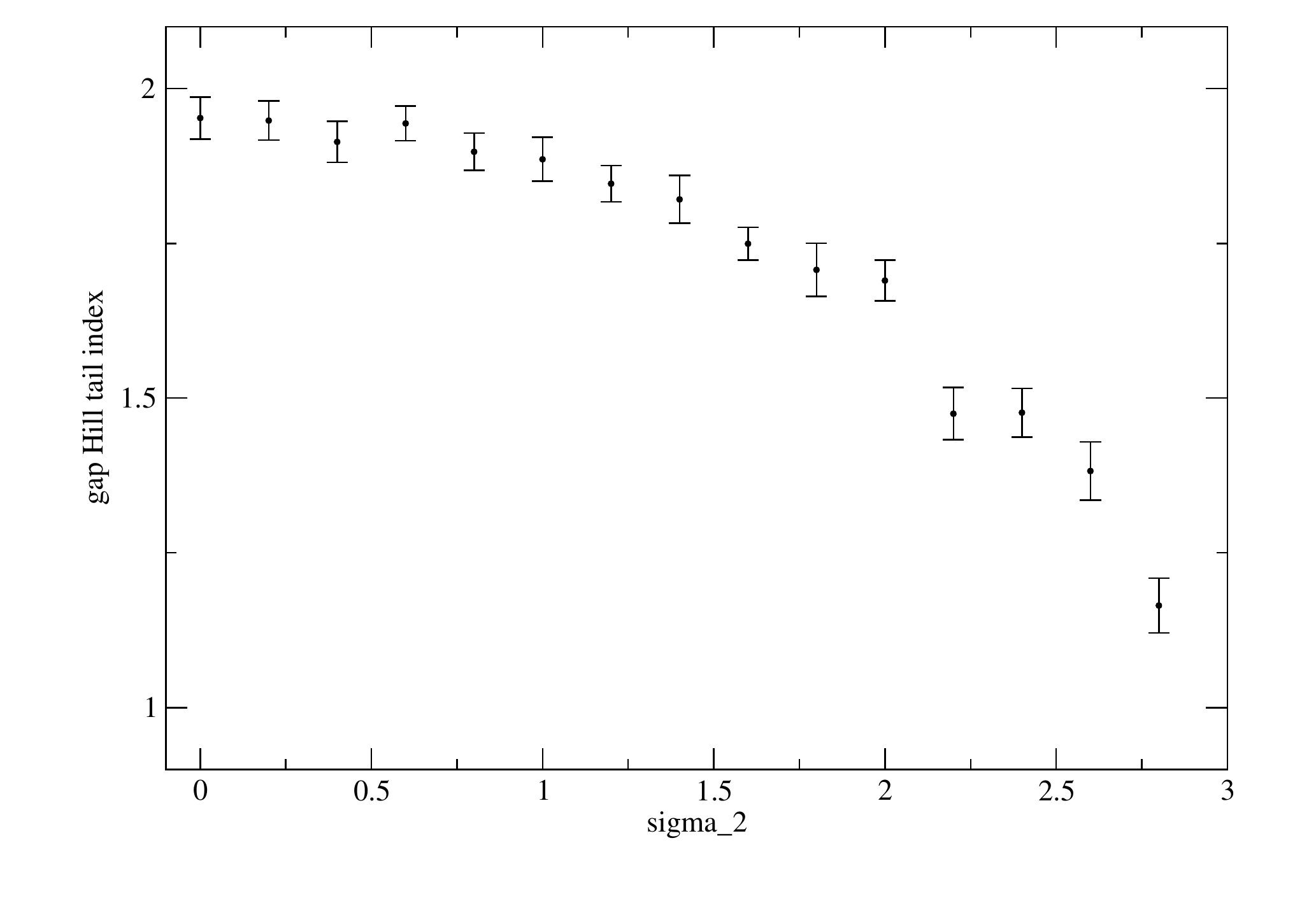}\includegraphics[width=6cm]{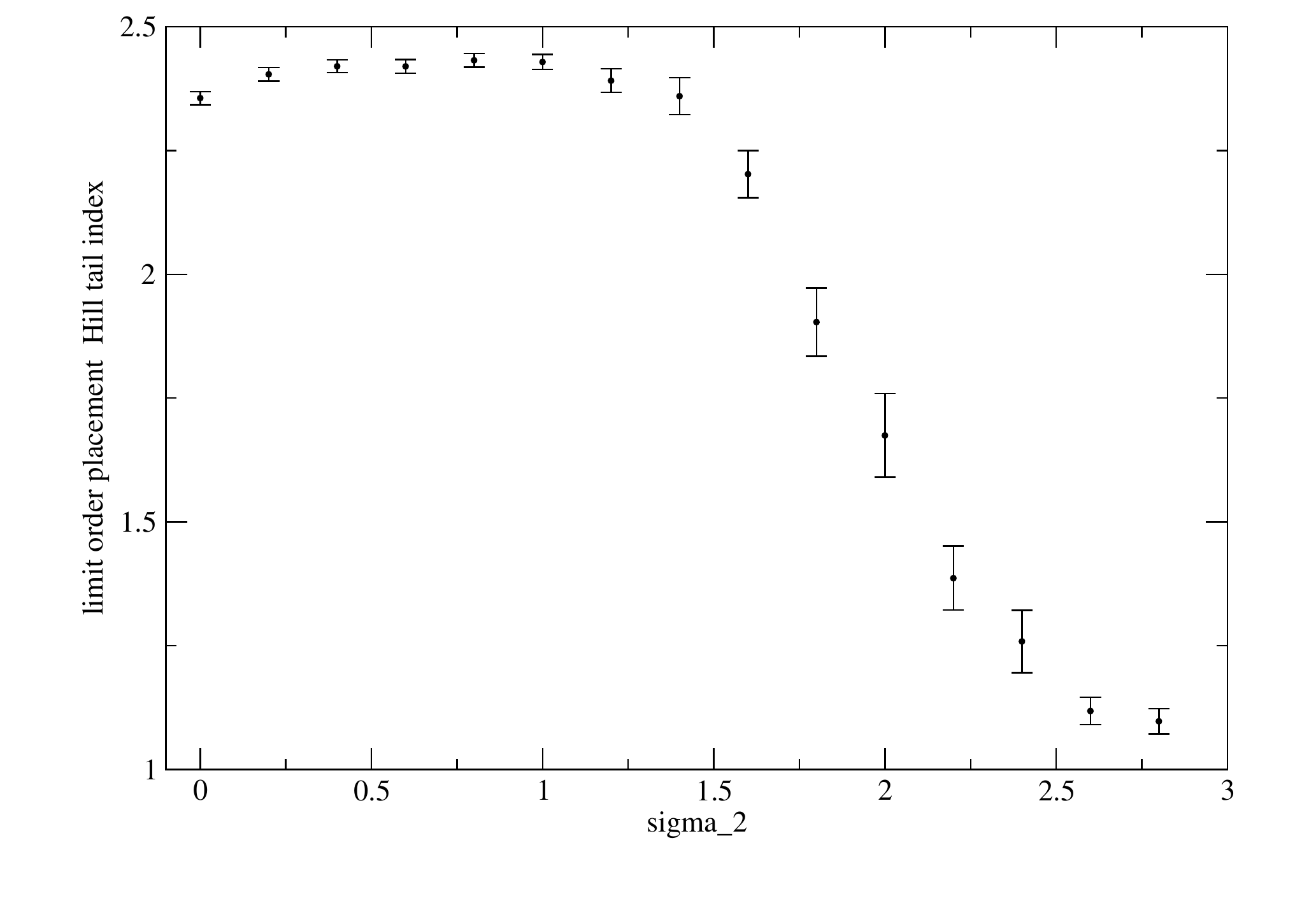}
\caption{(Left) Hill exponent of the first gap distribution for $\sigma_1 = 10.00$ and different values of $\sigma_2$. (Right) Hill exponent of the order placement distribution for for $\sigma_1 = 10.00$ and different values of $\sigma_2$. }
\label{figure9}
\end{figure}

Figure~\ref{figure9} shows the value of the Hill exponent for the
distribution of limit order placement from the best bid-ask (right).
The estimation of the Hill exponent gives values comparable with
those estimated empirically by Zovko and Farmer (2002) and Potters
and Bouchaud (2003). We then measured the distribution of gap sizes,
where the gaps are defined as the difference between the best price
(bid/ask) and the price at the next best quote (on each side
respectively). On the left side of Figure~\ref{figure9} we plot the
Hill exponent for the distribution of the first gap size and show
that it decreases with $\sigma_2$ (this observation, as discussed in
the next section, is important in understanding the origin of large
price fluctuations).

\begin{figure}[t]
\includegraphics[width=6cm]{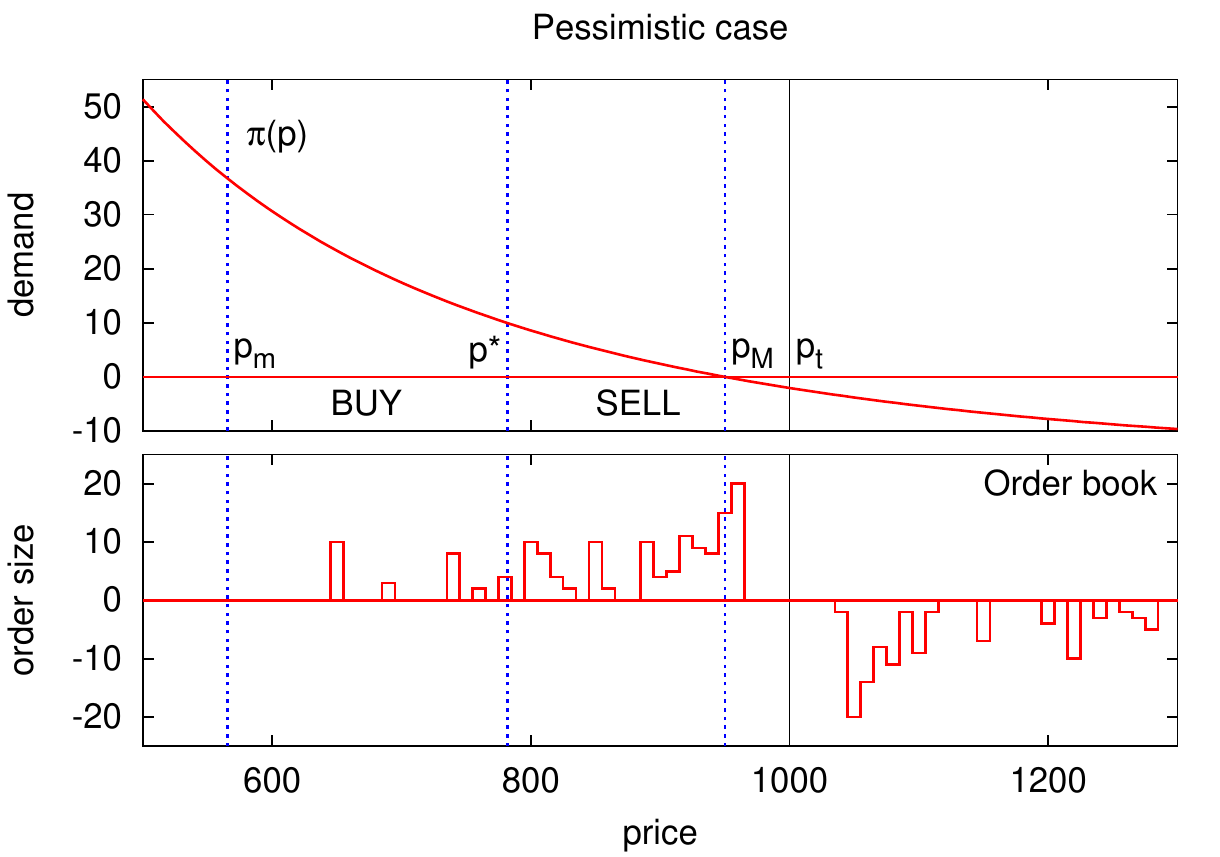}
\hspace{1cm}
\includegraphics[width=6cm]{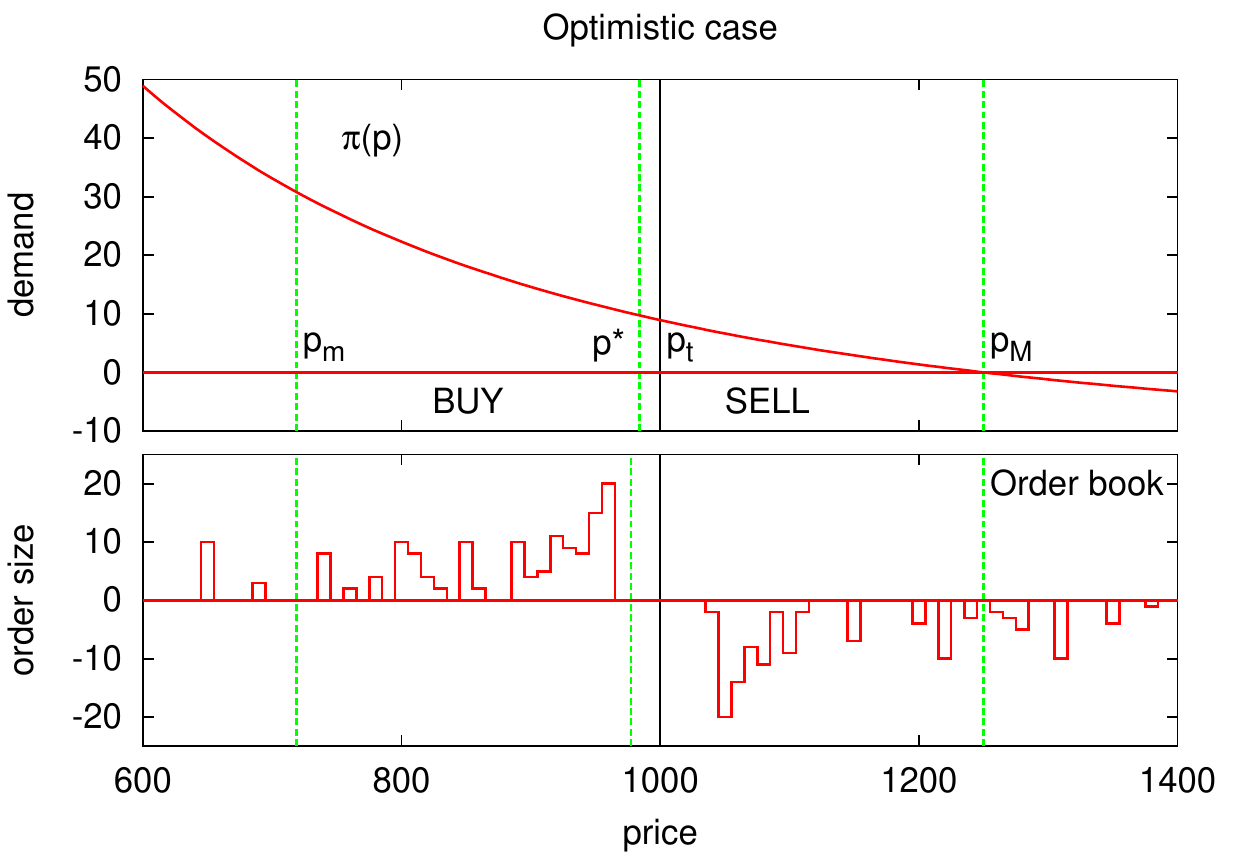} 
\caption{(Left) The role of the CARA demand function~(\ref{utility}) in the order placement decision when $p_M=\hat p^i_{t + \tau^i}<p_t$ and for a given
snapshot of the book. (Right) The role of the CARA demand function~(\ref{utility}) in the order placement decision when $p_M=\hat p^i_{t + \tau^i}>p_t$ and for a given snapshot of the book. The figures at the bottom show a  typical pattern of the limit order book at a given time in a simulation. Buy  orders placed in
the book are represented as positive volume, while sell orders placed in the book are represented as negative volume. The top figures are particular instances of Figure~\ref{figure}. The intersections determine the three different price levels $p_m$, $p^*$ and $p_M$ for the case of optimistic trader ($p_M> p_t$) (right) and pessimistic trader ($p_M<p_t$) (left).} 
\label{figure12}
\end{figure}

The origin of these large gaps can be explained as follows. As seen in Fig.~\ref{figure6} adding the chartist component induces traders to place orders far from the best bid and the best ask. The region where orders are placed is in fact four times larger than in the fundamentalist case. With the same number of orders spread over a broader region of the book, larger gaps between orders may arise both on the buy and sell side of the book. The chartist strategies also have a shorter time horizon $\tau^i$ that makes the book even more dynamic thus having stronger fluctuations in its shape and eventually creating even more gaps. Moreover, order submission differs for optimistic and pessimistic traders. An optimistic trader ($\hat{p}^i_{t+\tau^i}> p_t$)  can choose sell/buy orders at price levels $p \leq \hat p^i_{t + \tau^i}$ distributed symmetrically around the the current price $p_t$. A pessimistic trader ($\hat{p}^i_{t+\tau^i}< p_t$) instead typically chooses
orders at price levels $p< \hat p^i_{t + \tau^i}$ that are systematically below $p_t$. If the pessimistic trader chooses an order to sell, the order is very likely to be immediately executed as a market order (due to the fact that there is a high probability of finding a matching limit order to buy at a low price). On the
other hand, if the trader chooses an order to buy, it is likely to be a limit order placed at a large distance from the bid. Figure~\ref{figure12} shows this mechanism. It is this asymmetry between buy/sell orders that is responsible for generating larger gaps on the buy side of the book and thus the negative skewness shown in Fig.~\ref{figure4}.

\begin{figure}[t]
\includegraphics[width=6cm]{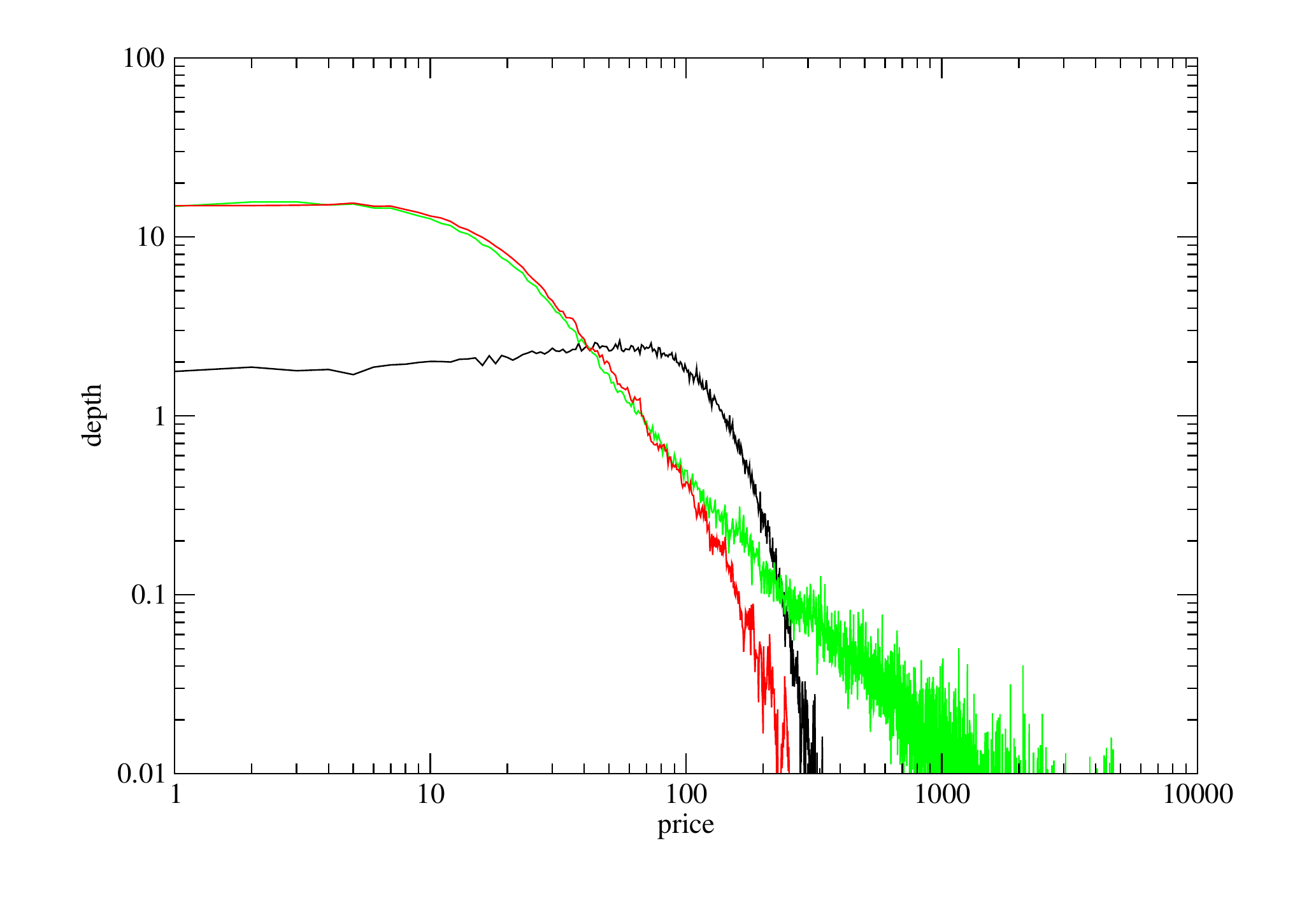}
\caption{The book shape (on log-log scale) with noise trading
(black), noise trading and fundamentalism (red), noise trading
fundamentalism and chartism (green). The book has a more realistic
shape with the depth higher toward the best price and fat tails
appearing as the chartist component $\sigma_2$ is activated. The
value of the parameters are the same as in Figure \ref{figure5}.}
\label{figure8}
\end{figure}

Figure~\ref{figure8} displays the book shape\footnote{While the
average shape of the book appears to be very smooth, a snapshot of
the book at a given time would show large gaps between orders and
several empty price ticks.} with noise trading (black), noise
trading and fundamentalism (red), noise trading fundamentalism and
chartism (green). The market depth is averaged over the entire
simulation period. Order placements are given as prices relative to
the midpoint between the best bid and best ask. We observe that the
book has an increasingly more realistic shape with the depth higher
toward the best price as the chartist component $\sigma_2$ is
activated. Chartist strategies also generate longer, power law,
tails in the distributions of orders in the book in qualitative
agreement with empirical findings (Bouchaud et al. (2002)).

In summary,  the risk aversion in the CARA utility
function~(\ref{utility}), has the effect of making noise traders
more impatient when they  sell than when they buy. Noise traders
prefer to sell immediately via market orders, and to buy by
submitting limit orders  at prices lower than the current quote. The
contribution of fundamentalist rules is that of reducing the
imbalance between buy and sell orders of both  types,  and of
concentrating the distribution of orders around the midpoint, thus
creating a more realistic shape of the book.  The effect of chartist
rules is that of widening the distribution of limit order
submissions farther away from the midpoint, thus  generating fatter
tails in the book and larger gaps between orders, particularly on
the buy side of the book.

\subsection{What generates large price fluctuations?}

\begin{figure}[t]
\includegraphics[width=6cm]{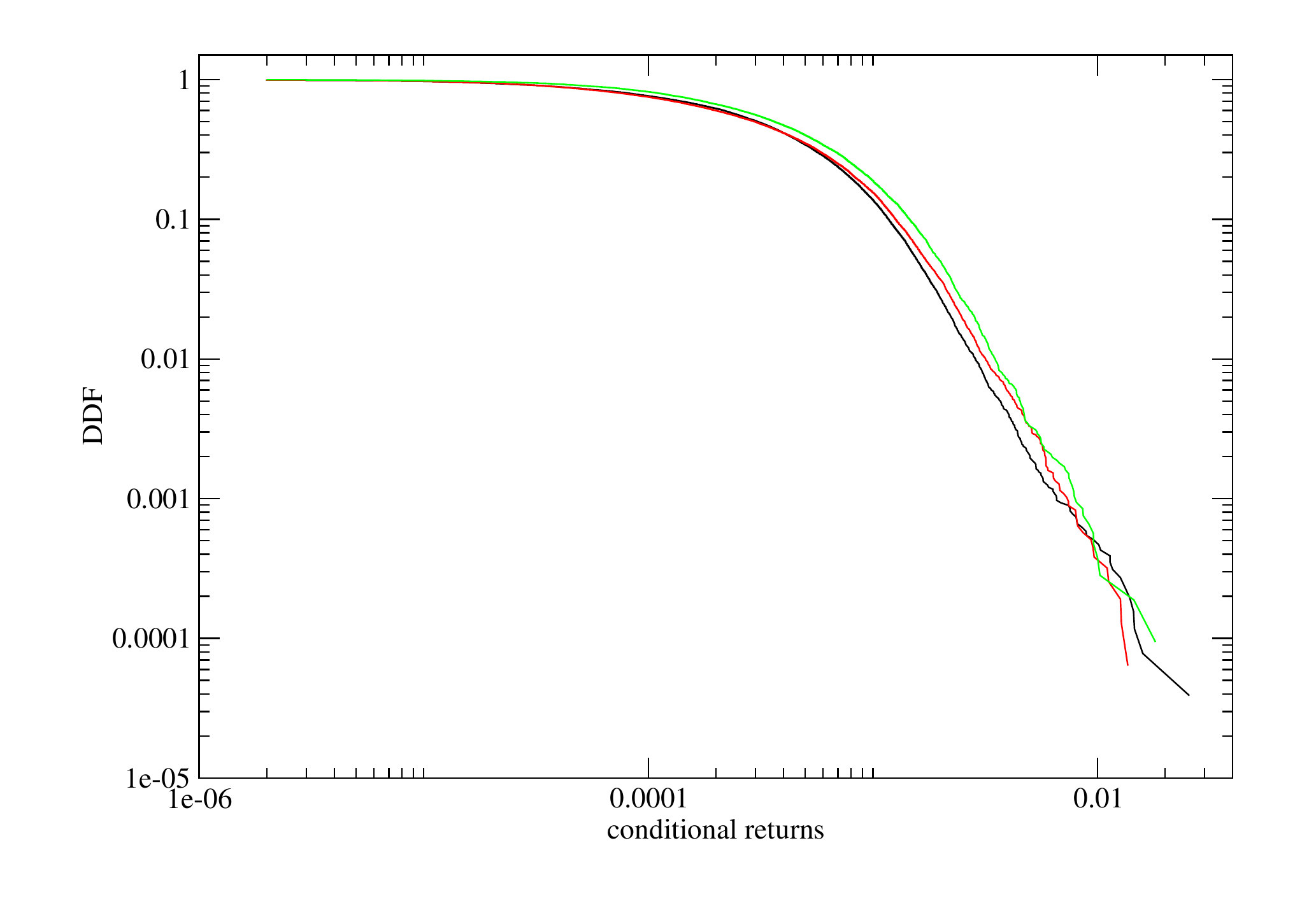}\includegraphics[width=6cm]{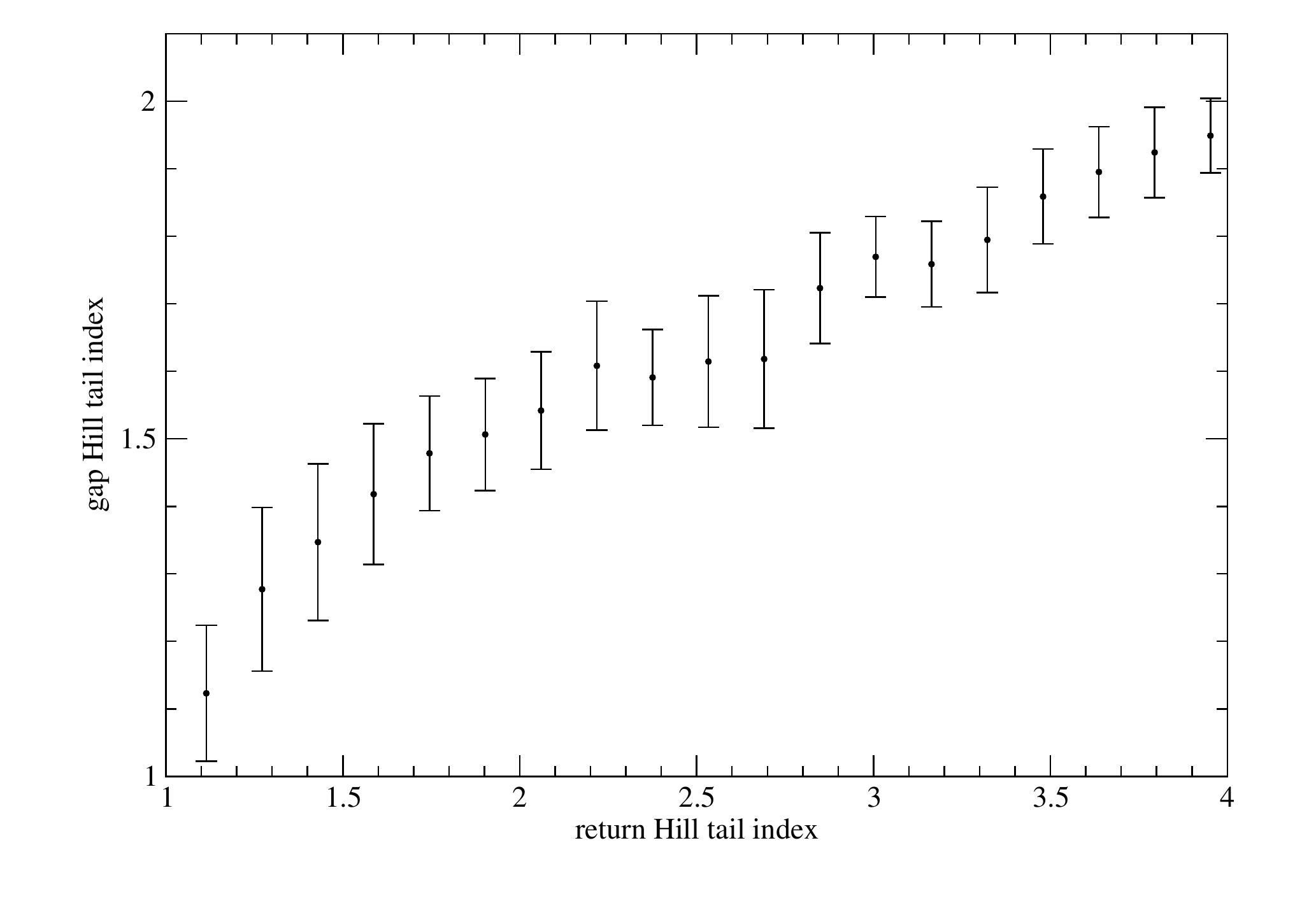}
\caption{(Left) Conditional returns for different trading volume: size $ <15$ (black), $15< $ size $ <30$ (red), size $ >30$ (green), for $\sigma_1= 10$ and $\sigma_2=2.0$. (Right) The tail index of the first gaps versus the tail index of returns for $\sigma_1= 10$ and different values of $\sigma_2$.}
\label{figure10}
\end{figure}

Next we address the fundamental question of what generates large
price returns, an issue that has been discussed by several authors
including Gabaix et al. (2003), Plerou et al. (2004), Farmer et al.
(2004), Farmer and Lillo (2004), and Lillo and Farmer (2005). We
compute the distribution of returns conditional on the size of
incoming market orders. To this end, we split orders into three
groups with approximately the same amount of orders in each. The
first group includes orders for buying or selling less than 15
stocks, the second group has orders of size between 15 and 30
stocks, and the last group orders of size larger than 30 stocks.
Figure~\ref{figure10} (left) displays the conditional distributions
of returns, given orders of different sizes, and shows the same
power law decay, indicating that large price fluctuations seem to be
rather insensitive to order size. In Fig.~\ref{figure10} (right) we
find instead an almost linear relationship between the Hill exponent
of returns distributions and the Hill exponent of the gap
distribution, indicating, as suggested by Farmer and Lillo (2004),
that the presence of large gaps at the best price is what drives
large price changes.

\section{Conclusion}
\label{conclude}

In this paper we have set up and analysed a model of a double
auction market. Each agent determines its demand through
maximisation of its expected utility of wealth, and bases its
expectation of future return on a fundamentalist, a chartist and a
noise trader component. Agents differ in risk aversion, investment
time horizon and the weight given to the three components of
expected return. We have thus extended a number of earlier models in
this literature which allowed agents to only place orders of unit
size, so that the model now incorporates feed back from the ongoing
evolution of the market.

We have used a number of statistical tools such as decumulative
distribution functions, the Hill tail index, and the rescaled range
statistic to analyse time series simulated by the model. We find
that the chartist component needs to be activated in order that the
generated returns exhibit many empirically observed features such as
volatility clustering, fat tails and long memory. We also find that
this approach describes fairly well a number of recently observed
stylized facts of double auction markets, such as the fat-tailed
distribution of limit orders placement from current bid/ask and the
fat-tailed distribution of orders stored in the order book.

Our paper also contributes to the debate on what generates large price changes in stock markets. Our simulations seem to confirm the picture proposed by Farmer and Lillo (2004), namely that large returns are driven by large gaps occurring in price levels adjacent to the best bid and best ask.

Future research could aim to enrich the economic framework of the model. For instance in place of allowing the weights on the fundamentalist, chartist and noise trader components in Eq.~(\ref{expect}) to the randomly selected they could be chosen on the basis of some fitness measure as in Brock and Hommes (1998). A further question of interest would be to see how the order book and order flow are affected if we assume agents have CRRA (constant relative risk aversion) utility functions, in which case their asset demands depend on their level of wealth. We know from Chiarella, Dieci and Gardini (2006) that in this type of modelling framework CRRA and
CARA utility functions lead to different types of dynamics.

\section*{Acknowledgments}

JP acknowledges kind hospitality of City University during his stay with grant 2004-BE-00314 from Ag\`encia de Gesti\'o d'Ajuts Universitaris i de Recerca and also financial support by Direcci\'on General de Investigaci\'on under contract No. FIS2006-05204. GI acknowledges generous hospitality and financial support from UTS during her stay in Sydney in August 2003. Carl Chiarella acknowledges financial support from Australian Research Council grant DP0450526. We would like to thank (Tony) Xuezhong He for helpful comments.

\section{References}

\noindent Bottazzi, G., Dosi, G., Rebesco, I., 2005. Institutional Achitectures and Behavioural Ecologies in the Dynamics of Financial Markets: a Preliminary Investigation. Journal of Mathematical Economics 41, 197-228.

\noindent Bouchaud, J.-P., M$\acute{e}$zard, M., Potters, M., 2002. Statistical properties of stock order books: empirical results and models. Quantitative Finance 2, 251-256.

\noindent Bouchaud, J.-P., Kockelkoren, J., Potters, M., 2006. Random Walks, liquidity molasses and critical response in financial markets. Quantitative Finance 6, 115-123.

\noindent Bouchaud, J.-P., Gefen, Y., Potters, M., Wyart, M., 2004. Fluctuation and response in Financial Markets: the subtle nature of Random price changes. Quantitative Finance 4, 176-190.

\noindent Brock, W., Hommes, C., 1998. Heterogeneous beliefs and routes to chaos in a simple asset pricing model. Journal of Economics Dynamics and Control 22, 1235-1274.

\noindent Challet, D., Stinchcombe, R., 2001. Analyzing and Modelling 1+1d markets. Physica A 300, 285-299.

\noindent Chiarella, C., Dieci, R., Gardini, L., 2006. Asset price and wealth dynamics in a financial market with heterogeneous agents. Journal of Economic Dynamics and Control 30, 1755-1786.

\noindent Chiarella, C., Iori, G., 2002. A simple microstructure model of double auction markets. Quantitative Finance 2, 346-353.

\noindent Consiglio, A., Lacagnina, V., Russino, A., 2005. A Simulation Analysis of the Microstructure of an Order Driven Financial Market with Multiple Securities and Portfolio Choices. Quantitative Finance 5, 71-88.

\noindent Farmer, J.D., Gillemot, L., Lillo, F., Mike, S., Sen, A., 2004. What really causes large price changes?. Quantitative Finance 4, 383-397.

\noindent Farmer, J.D., Lillo, F., 2004. On the origin of power laws in financial markets. Quantitative Finance 4, C7-C10.

\noindent Daniels, M.G., Farmer, J.D., Gillemot, L., Iori, G., Smith, E., 2003. Quantitative model of price diffusion and market friction based on trading as a mechanistic random process. Physical Review Letters 90, 108102.

\noindent Gabaix, X., Gopikrishnan, P., Plerou, V., Stanley, H. E., 2003. A theory of power-low distributions in financial market fluctuations. Nature 423, 267-270.

\noindent Gillemot, L., Farmer, J.D., Lillo, F., 2006. There's More to Volatility than Volume, Quantitative Finance 6, 371-384.

\noindent Gil-Bazo, J., Moreno, D., Tapia, M., 2007. Price Dynamics, Informational Efficiency and Wealth Distribution in Continuous Double Auction Markets. Computational Intelligence 23, 176-196.

\noindent Hill, B.M., 1975. A simple general approach to inference about the tail of a distribution. Annals of Statistics 3, 1163-1173.

\noindent Hommes, C.H., 2001. Financial markets as nonlinear adaptative evolutionary systems. Quantitative Finance 1, 149-167.

\noindent Hurst, H., 1951. Long Term Storage Capacity of Reservoirs. Transactions of the American Society of Civil Engineers 116, 770-799.

\noindent Kirman, A., Teyssiere, G., 2002. Microeconomic Models for Long Memory in the Volatility of Financial Time Series. Studies in Nonlinear Dynamics and Econometrics 5, 281-302.

\noindent Li Calzi, M., Pellizzari, P., 2003. Fundamentalists clashing over the book: a study of order-driven stock markets. Quantitative Finance 3, 470-480.

\noindent Lillo, F., Farmer, J.D., Mantegna, R.N., 2003. Single Curve Collapse of the Price Impact Function for the New York Stock Exchange. Nature 421, 129-130.

\noindent Lillo, F., Farmer, J.D., 2004. The Long Memory of the Efficient Market. Studies in Nonlinear Dynamics \& Econometrics 8, 1-33.

\noindent Lillo, F., Farmer, J.D., 2005. The Key Role of Liquidity Fluctuations in Determining Large Price Fluctuations. Fluctuations and Noise Letters 5, L209-L216.

\noindent Lillo, F., Mike, S., Farmer, J.D., 2005. Theory for Long Memory in Supply and Demand. Physical Review E 7106 287-297.

\noindent Lo, A., 1991. Long-Term Memory in Stock Market Prices. Econometrica 59, 1279-1313.

\noindent Luckock, H., 2003. A steady-state model of the continuous double auction. Quantitative Finance 3, 385-404.

\noindent Lux, T. , 2001. The limiting extremal behaviour of speculative returns: an analysis of intra-daily data from the Frankfurt Stock Exchange. Applied Financial Economics 11, 299-315.

\noindent Mandelbrot, B., 1972. Statistical Methodology for Non-periodic Cycles: From the Covariance to R/S analysis. Annals of Economic and Social Mesurements 1, 259-290.

\noindent Plerou, V., Gopikrishnan, P., Gabaix, X., Stanley, H.E., 2004. On the origin of power-law fluctuations in stock prices. Quantitative Finance 4, C11-C15.

\noindent Potters, M., Bouchaud, J.-P., 2003. More statistical properties of stock order books and price impact. Physica A 324, 133-140.

\noindent Potters, M., Bouchaud, J.-P., 2006. Trend followers lose more often than they gain. Wilmott Magazine Jan 2006.

\noindent Raberto, M., Cincotti, S., Focardi, S.M., Marchesi, M., 2001. Agent-based simulation of a financial market. Physica A 299, 320-328.

\noindent Slanina, F., 2007. Critical comparison of several order-book models for stock-market fluctuations. The European Physical Journal B, forthcoming.

\noindent Weber, P., Rosenow, B., 2005. Order book approach to price impact. Quantitative Finance 5, 357-364.

\noindent Zovko, I.I., Farmer, J.D., 2002. The power of patience: A behavioral regularity in limit order placement. Quantitative Finance 2, 387-392.

\begin{appendix}

\section{Demand function and trading mechanism\label{demand}}

At the beginning of each trading timestep, the agent constructs its individual demand function and determines the amount of wealth it would like to invest in the risky asset for any possible level of the notional transaction price level $p$. The residual wealth is invested in a riskless asset with zero interest rate.

For simplicity, in this Appendix we drop the agent superscript $i$ from all variables. The following procedure for the determination of individual demand functions is understood to apply to every agent $i\in \{1,\cdots, I\}$. We first assume that the agent's decision is taken in terms of a CARA (Constant Absolute Risk Aversion) class as given by Eq.~(\ref{utility})
\begin{equation}
U(W, \alpha)=-\exp(-\alpha W),
\label{utility2}
\end{equation}
where $\alpha$ is the risk aversion of the agent. The wealth $W_{t}$
at a given time step is given by a cash amount $C_{t}$ and the
quantity $S_{t}$ of the risky asset whose value is $p_{t}$. That is:
$W_{t}=C_{t}+S_{t}p_{t}$.

At time $t$ the agent places  an order of size $S$ with price level
$p$. The agent takes these two decisions about size and price level
based on its own expected price of the stock at a time horizon
$\tau$ (cf. Eqs.~(\ref{pexpect}) and~(\ref{tau})) and the
maximization of its own expected utility function at horizon $\tau$.
Assuming the CARA class utility function~(\ref{utility2}), the
invertor's expected utility at time $t+\tau$ is given by
\begin{equation}
\hat{U}_{t+\tau}=\mathbb{E}_{t}\left[-\exp{(-\alpha W_{t+\tau})}\right],
\label{w1}
\end{equation}
where $\mathbb{E}_{t}\left[\cdot \right]$ stands for the expected mean conditional to the information available at time $t$, right before placing the order, with current price level $p_t$. The wealth
at time $t+\tau$ (assuming the order is executed at the price $p$ before $t+\tau$) is given by
$$
W_{t+\tau}=W_t+S_t p \rho_{t+\tau},
$$
where $\rho_{t+\tau}=p_{t+\tau}/p-1$ is the return from $t$ to
$t+\tau$, and the wealth is proportional to $\rho_{t+\tau}$. As a
zero order approximation, the agent's expectation for future returns
are taken to be Gaussian where future return is assumed to be
$\rho_{t+\tau}=p_{t+\tau}/p-1\simeq\ln(p_{t+\tau}/p)$. This
assumption allows us to write Eq.~(\ref{w1}) as
%\begin{equation}
$$
\hat{U}_{t+\tau}=-\exp{(-\alpha
\mathbb{E}_{t}\left[W_{t+\tau}\right]+ \alpha^2 V_{t}\left[W_{t+\tau}\right]/2)},
%\label{w2}
%\end{equation}
$$
where $V_{t}\left[\cdot\right]$ stands for the conditional variance. Since $\mathbb{E}_{t}\left[W_{t+\tau}\right]=W_t+S_tp \mathbb{E}_{t}\left[\rho_{t+\tau}\right]$ and $V_{t}\left[W_{t+\tau}\right]=S_t^2p^2V_{t}\left[\rho_{t+\tau}\right]$,
we get
\begin{equation}
\hat{U}_{t+\tau}=U(W_t)\exp{\left(-\alpha S_t p\mathbb{E}_{t}\left[\rho_{t+\tau}\right]+\alpha^2
S_t^2p^2V_{t}\left[\rho_{t+\tau}\right]/2\right)}.
\label{w3}
\end{equation}
At time $t+\tau$ the agent expects that the price will be
$\hat{p}_{t+\tau}$ (cf. Eq.~(\ref{pexpect})) and the expected return
thus reads
$\mathbb{E}_{t}\left[\rho_{t+\tau}\right]=\ln(\hat{p}_{t+\tau}/p)$.
However, our model does not specify any expectation mechanism for
the return variance. For this reason we take
$V_{t}\left[\rho_{t+\tau}\right]=V_t$ as the historical one
calculated over a certain time window as shown in
Section~\ref{TheModel} in Eq.~(\ref{sigma}).

Differentiating the expected utility function~(\ref{w3}) with respect to $S_t$ gives
$$
%\begin{equation}
\frac{d}{dS_t}\hat{U}_{t+\tau}=-\hat{U}_{t+\tau}\alpha p\left(\ln(\hat{p}_{t+\tau}/p)-\alpha S_tp V_t\right).
%\label{w4}
%\end{equation}
$$
Setting the expression above to zero we determine the optimal amount
of stocks, $S^*_t=\pi(p)$ the agent wishes to hold in its portfolio
at time $t$ for a given price level $p$, based on expectations at
time horizon $\tau$, namely
%\begin{equation}
$$
\pi(p)=\frac{\ln(\hat{p}_{t+\tau}/p)}{\alpha V_tp},
%\label{w5}
%\end{equation}
$$
which coincides with Eq.~(\ref{equation8}).

Before going back to the main text, it is worth mentioning that
Bottazzi et. al (2005) give a similar analysis but with the
hypothesis that the expected utility function reads
\begin{equation}
\hat{U}_{t+\tau}=\mathbb{E}_{t}\left[W_{t+\tau}\right]-\frac12 \alpha V_{t}\left[W_{t+\tau}\right].
\label{w6}
\end{equation}
This is simply a function depending linearly on the expected return and its variance (see also Brock and Hommes (1998), Hommes (2001) and Kirman and Tessyiere (2002). By neglecting higher moments one is in practice assuming Gaussianity in the agent's expectations, as we do.

\section{Hill estimator\label{Hill}}

In order to quantify the fat tails of our distribution of returns we estimate the Hill tail index (Hill (1975)). The Hill estimator is a maximum likelihood estimator of the parameter $\beta$ of the Pareto law $F(x)\sim 1/x^\beta$ for large $x$, where $F(x)$ denotes the return pdf. Because of its simplicity the Hill estimator has become the standard tool in most studies of tail behaviour of economic data.

To compute the Hill tail index, the sample elements are put in descending order: $x_n\geq x_{n-1}\geq\cdot\cdot\cdot \geq x_{n-k}\geq \cdot\cdot\cdot\geq x_1$ where $n$ is the length of our data sample and $k$ is precisely the number of observations located in the tail of our distribution. Hence, the Hill estimator obtains the inverse of $\beta$ as
\begin{equation}
\hat{\gamma}_H(n,k)=\frac{1}{\hat{\beta}_H\left(n,k\right)}=\frac{1}{k}\sum_{i=1}^k\left[\ln x_{n-i+1}-\ln x_{n-k}\right].
\label{hill}
\end{equation}
The main difficulty with this procedure is to choose the optimal threshold $k^\star$. If we take $k$ too small we may have too few statistics, while if we include too large a data set inside the tail then the estimator increases because of contamination with entries from more central parts of the distribution. There are many
techniques aimed at finding this optimal cut-off (see for instance Lux (2001)). All of them agree that the optimal $k^\star$ is defined as the number of order statistics minimizing the mean squared error
of $\hat{\gamma}_H(n,k)$ defined in Eq.~(\ref{hill}). However, the optimal sample fraction is not easy to infer from this assertion and one of the most common ways to perform the estimation in the literature is by a bootstrapping technique (again see Lux (2001)).

In our case, we have evaluated the Hill tail index for a fixed threshold taking 5\% of the data. While this choice does not necessarily provide the optimal cut-off, this is the value commonly taken in the literature (see e.g. Lux (2001)). We have carried out the Hill estimator procedure to estimate the right and the left tail
exponents of both the return and also the absolute return time series.

The estimated tail index $\hat{\gamma}=1/\hat{\beta}$ given by Eq.~(\ref{hill}) is obtained for every simulation we ran. It can be shown that $(\hat{\gamma}-\gamma)\sqrt{k}$ is asymptotically normal with zero mean and variance $\gamma^2$ where $\gamma=1/\beta$ (see Lux (2001) for more details). The confidence level of being inside the interval of one unit of mean squared error is 67\%, being inside the interval of two units of mean squared error is 95\%, and being inside the interval of three units of mean squared error is 99.7\%.

All these properties allow us to give the error of our measurement and its degree of confidence. The output of our simulations gives the average of the Hill tail index over all the path simulations and its standard deviation. The estimation should converge to the "true" Hill tail index in accordance to the Central Limit Theorem. Hence, as the number of simulations increases
$$
\langle\hat{\gamma}\rangle \equiv \frac{1}{N}\sum_{i=1}^N \hat{\gamma}_i\rightarrow \gamma,
$$
and
$$
\sigma_{\gamma}^2 =\frac{1}{N(N-1)}\sum_{i=1}^N \left(\hat{\gamma}_i-\langle\hat{\gamma}\rangle\right)^2,
$$
where $N$ is the total number of simulations and $i$ is the index
for each simulation. So that we obtain as the confidence band for
$\gamma$,
$$
\langle\hat{\gamma}\rangle\pm\frac{\sigma_{\gamma}}{\sqrt{N}},
$$
and if one wants to focus on the tail index this reads
$$
\frac{1}{\langle\hat{\gamma}\rangle}\pm\frac{\sigma_{\gamma}}{\langle\hat{\gamma}\rangle^2\sqrt{N}}.
$$

\end{appendix}

\end{document}